\documentclass[sigconf]{acmart}
\usepackage{graphicx} 
\usepackage[normalem]{ulem}
\usepackage{algorithm}
\usepackage{algpseudocode}%
\usepackage{array}
\usepackage{booktabs}
\usepackage{subcaption}
\usepackage{multirow}
\usepackage{amsfonts}
\usepackage{todonotes}

\AtBeginDocument{%
  \providecommand\BibTeX{{%
    \normalfont B\kern-0.5em{\scshape i\kern-0.25em b}\kern-0.8em\TeX}}}

\setcopyright{acmcopyright}
\copyrightyear{2018}
\acmYear{2018}
\acmDOI{10.1145/1122445.1122456}

\acmConference[Woodstock '18]{Woodstock '18: ACM Symposium on Neural
  Gaze Detection}{June 03--05, 2018}{Woodstock, NY}
\acmBooktitle{Woodstock '18: ACM Symposium on Neural Gaze Detection,
  June 03--05, 2018, Woodstock, NY}
\acmPrice{15.00}
\acmISBN{978-1-4503-XXXX-X/18/06}

\begin{document}

\title{OPAD: An Optimized Policy-based Active Learning Framework for Document Content Analysis}

\author{Sumit Shekhar}
\email{sushekha@adobe.com}
\affiliation{%
  \institution{Adobe Research}
  \country{India}}
  
\author{Bhanu Prakash Reddy Guda}
\email{guda@adobe.com}
\affiliation{%
  \institution{Adobe Research}
  \country{India}}
  
\author{Ashutosh Chaubey}
\email{achaubey@cs.iitr.ac.in}
\affiliation{%
  \institution{IIT Roorkee}
  \country{India}}
  
\author{Ishan Jindal}
\email{ijindal@ec.iitr.ac.in}
\affiliation{%
  \institution{IIT Roorkee}
  \country{India}}
  
\author{Avneet Jain}
\email{ajain1@ee.iitr.ac.in}
\affiliation{%
  \institution{IIT Roorkee}
  \country{India}}


\begin{abstract}
  Documents are central to many business systems, and include forms, reports, contracts, invoices or purchase orders. The information in documents is typically in natural language, but can be organized in various layouts and formats. There have been recent spurt of interest in understanding document content with novel deep learning architectures. However, document understanding tasks need dense information annotations, which are costly to scale and generalize. Several active learning techniques have been proposed to reduce the overall budget of annotation while maintaining the performance of the underlying deep learning model. However, most of these techniques work only for classification problems. But content detection is a more complex task, and has been scarcely explored in active learning literature. In this paper, we propose \textit{OPAD}, a novel framework using reinforcement policy for active learning in content detection tasks for documents. The proposed framework learns the acquisition function to decide the samples to be selected while optimizing performance metrics that the tasks typically have. Furthermore, we extend to weak labelling scenarios to further reduce the cost of annotation significantly. We propose novel rewards to account for class imbalance and user feedback in the annotation interface, to improve the active learning method. We show superior performance of the proposed \textit{OPAD} framework for active learning for various tasks related to document understanding like layout parsing, object detection and named entity recognition. Ablation studies for human feedback and class imbalance rewards are presented, along with a comparison of annotation times for different approaches.
\end{abstract}


\begin{teaserfigure}
    \centering
    \includegraphics[width=1.0\linewidth]{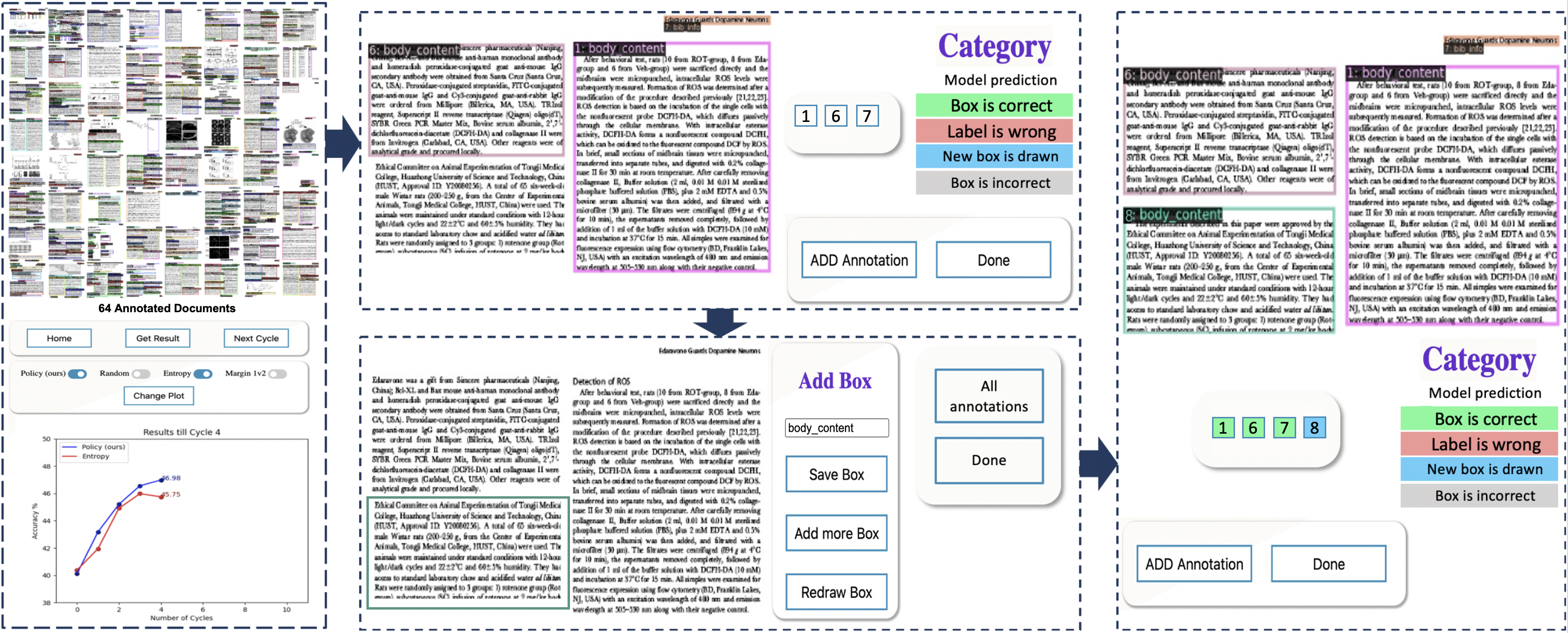}
    \label{fig:annot_tool}
    \caption{The proposed active learning-based interface, OPAD, enables intelligent annotation functionalities like optimized selection of documents for layout detection task, annotating instances and text with boxes (strong labelling) or verification and correction of annotations (for weak labelling scenarios). \textit{Document is cropped for better visualization}}
    \label{fig:annotationinterface}
\end{teaserfigure}
\maketitle

\section{Introduction}
Documents are a key part of several business processes, which can include reports, business contracts, forms, agreements, etc. Extracting data from documents through deep networks have recently started gaining attention. These tasks include document page segmentation, entity extraction or classification. Fueled by the availability of both labeled and unlabeled data, and advances in the computation infrastructure, recently, a number of deep learning models have been proposed for modeling complex tasks~\cite{he2015Deep,ren2015FasterRcnn,devlin2019Bert}. The promising results from this research direction motivated development of several deep learning models which show significant performance improvements on these tasks when trained on a large amount of labelled data~\cite{oliveira2018dhsegment,yang2016joint,xu2020layoutlm}. However, deployment of these models requires considerable \textit{effort} and \textit{cost} to annotate unlabelled data especially for document tasks because of requirements for dense annotations, e.g. annotating page structures with components like \textit{title}, \textit{table}, \textit{figures} or \textit{references}. Thus, there is a need to explore methods to optimize annotation budgets to accelerate the development of document analysis models.

Several approaches have been proposed in the domain of semi-supervised learning~\cite{yang2021survey}, unsupervised learning~\cite{wilson2020survey}, few-shot learning~\cite{wang2020generalizing}, active learning~\cite{settles2009Active} etc$\dots$ to overcome the limitation of availability of labeled data. Each of these approaches have their own objectives incorporated in either modeling or data annotation or both for achieving superior performances in a limited annotated data setup. Among these, our motivation for using active learning is two-folds: (1) active learning bridges the gap in the model by querying samples in the data space, for which the model does not have enough information~\cite{settles2009Active}, (2) the active learning approaches seek to learn higher accuracy models within a given annotation cost, through optimizing data acquisition, which align well with our objective of optimizing annotation costs. Recent methods for \textit{pool-based active learning} scenario, the query for annotations selects a subset batch of data samples for the oracle (\textit{i.e.} the annotator). Pool or batch-based active learning methods are more scalable than querying single data sample per learning cycle~\cite{guo2007discriminative}. Most of the active learning work \cite{settles2009Active,aggarwal2014Active} formulate acquisition functions as information theoretic uncertainty estimates. While uncertainty-based methods work well for tasks like classification \cite{wang2014Heuristic,gal2017BayesianAL}, where a single annotation is required per data sample, generalizations to document tasks such as page segmentation and named entity recognition, which require multiple annotations per selected data sample, have been scarcely explored. This is because methods to aggregate uncertainties over various entities present in a data sample are not well developed \cite{roy2018DetectionHeuristicAL,brust2019DetectionHeuristicAL}. Recent techniques have been proposed to obtain a better acquisition function for active learning in these tasks \cite{liu2018ImmitationAL,liu2019PersonReIdAL}. However, these methods assume highly task-specific heuristics, and hence can not be generalized across different content detection scenarios. 

In addition to active learning, in particular for dense annotation tasks in documents, weak learning can be an effective approach to reduce annotator's efforts \cite{papadopoulos2016NoBBoxesHumanVerification,wang2017CEAL,papadopoulos2017ClickSupervision}. When there are multiple entities to be annotated in a data sample, weak learning reduces the annotation effort, either by providing faster variations of annotation techniques \cite{papadopoulos2017ClickSupervision} or simply asking the annotator to verify the model predictions \cite{papadopoulos2016NoBBoxesHumanVerification}. However, there are very few works \cite{desai2019AnAS,brust2020ALWeakSupervision} that combine weak learning with active learning. Furthermore, to the best of our knowledge, none of the works takes advantage of the annotator feedback (e.g. from annotator's corrections of detected instance boundaries) during an active learning cycle.

In this work, we propose a policy-based active learning approach, taking into account the complexities of aggregating model uncertainties in the selection of samples to be labelled. We model the task of active learning as a Markov decision process (MDP) and learn an optimal acquisition function using deep $Q$-learning \cite{mnih2013DeepQLearning}. While several works rely on reinforcement learning for learning an optimal acquisition function \cite{liu2018ImmitationAL,liu2019PersonReIdAL,haussmann2019CLassRLAL,casanova2020Reinforced}, they assume task-specific representations of states and actions and hence are not generalizable across tasks. We further show that the proposed method can be combined with weak labelling, reducing the cost of annotation compared to strong labelling. Moreover, we incorporate class imbalance and human feedback signals into the design of MDP using suitable reward functions to further improve the performance of our approach.

To summarize, the major contributions of our work are as follows:
\begin{itemize}
    \item We propose a policy-based task-agnostic active learning approach for complex content detection tasks, layout detection and named entity recognition in documents.
    \item We report that the proposed approach is generalizable, through demonstrating the performance of our active learning setup on varied detection tasks.
    \item We investigate the effectiveness of incorporating class balance and human feedback rewards in improving the active learning policy.
    \item We demonstrate the advantage of the proposed approach in reducing the costs of annotation in aforementioned complex detection tasks.
\end{itemize}

Throughout the remainder of the paper, we explain the proposed concepts, models, configurations, and discussions from the perspective of the layout and object detection, and named entity recognition tasks.

\section{Related Work}
Document content analysis has been studied extensively along several dimensions such as document classification ( image~\cite{xu2020layoutlm,xu2020layoutlmv2} or text~\cite{adhikari2019docbert,pappagari2019hierarchical} or both~\cite{jain2019multimodal,audebert2019multimodal}), named entity recognition in documents~\cite{yang2016joint,luo2018attention}, content segmentation~\cite{oliveira2018dhsegment,gruning2019two}, document retrieval~\cite{sugathadasa2018legal,choudhary2020document,trabelsi2021neural}, layout analysis~\cite{binmakhashen2019document} among many others. The availability of large scale labeled datasets of documents~\cite{lewis2006building,harley2015evaluation,li2020docbank, Tkaczyk2014GROTOAP2T,zhong2019publaynet} led to the advent of several state-of-the-art deep learning models which have significantly improved these tasks in a large scale data setup. However, to the best of our knowledge, there is very limited amount of literature which uses active learning to optimize data annotation cost in a low resource setting, specifically for document analysis tasks~\cite{godbole2004document,bouguelia2013stream}. Therefore, in this section, we discuss about works that deal with general active learning policies, and active learning in a couple of related well studied domains, image classification, object detection and named entity recognition. 

Active learning selects data samples with high uncertainty in the model prediction, which can provide more information to the underlying model. Different works have proposed different ways to compute model uncertainty~\cite{settles2009Active}. While some methods depend on information theory for designing acquisition functions \cite{houlsby2012BALD,wang2014Heuristic,gal2017BayesianAL}, others rely on alternative ways to approximate model uncertainty \cite{freytag2014ALModelOutputChange,ducoffe2018Adversarial}. Yoo \textit{et al}~\cite{yoo2019LearningLossAL} add a light-weight loss prediction module to the prediction model to predict the loss for the unlabelled samples, and use that as an uncertainty measure. Mayer \textit{et al}~\cite{Mayer2020AdversarialSF} use uncertainty measure to find the optimal sample and query the data sample closest to the optimal sample. 

For complex tasks such as object detection and named entity recognition, recent works \cite{roy2018DetectionHeuristicAL,brust2019DetectionHeuristicAL,shen2017NERHeuristicAL} have been proposed to use uncertainty scores for the acquisition of samples. Most of these methods rely on aggregating the uncertainties of various entities within a data sample using max, sum or average functions \cite{roy2018DetectionHeuristicAL,brust2019DetectionHeuristicAL}. Aghdam \textit{et al}~\cite{aghdam2019PedestrianAL} proposed a novel approach combining pixel-level scores to obtain an image-level score for doing active learning for the task of pedestrian detection task. 

Several works have been proposed to incorporate reinforcement learning to learn an optimal acquisition function for active learning. The objective of these approaches is to model the active learning process into a Markov decision process through defining and designing suitable representations for states, actions, and rewards \cite{fang2017NlpRLAL,liu2018MtRLAL,haussmann2019CLassRLAL}. Liu \textit{et al}~\cite{liu2018ImmitationAL} proposed an imitation learning approach for active learning in tasks related to natural language processing, relying on an algorithmic expert to find an optimal acquisition function. We differ from the work of Casanova \textit{et al}~\cite{casanova2020Reinforced} on using reinforced active learning approach for image segmentation, in terms of the generalize-ability of our approach on various tasks. We also report the effectiveness of using weak learning on top of policy-based active learning in consuming the budget with maximum efficiency.

\section{Proposed OPAD Framework}
In this section, we describe the proposed \textbf{O}ptimized \textbf{P}olicy-based \textbf{A}ctive Learning Framework for \textbf{D}ocument Content Analysis, \textit{OPAD}. Figure~\ref{fig:annotationinterface} shows the interface for \textit{OPAD}, which enables various scenarios of detection tasks for human annotators. The underlying algorithm for \textit{OPAD} is a Deep Query Network (DQN)-based reinforcement learning policy, optimized for data sample selection based on the performance metrics for the task. \textit{OPAD} has two stages - policy training stage and deployment stage. In the policy training stage, \textit{OPAD} is trained using simulated active learning cycles to maximize performance on a validation set. While deploying, the trained policy is used to make online batch selection for annotation. The overall formulation for OPAD is described below.

\subsection{Formulation}
The underlying objective for policy training in \textit{OPAD} is to perform an iterative selection of the samples from an unlabelled pool, $\mathscr{X}_u$, which would maximally increase the performance of the model being trained, $\Theta$ until the annotation budget, $\mathbb{B}$ is consumed. In each active learning cycle, the policy DQN $\Pi$ \cite{mnih2013DeepQLearning} selects a batch of $n_{cycle}$ samples, which are labelled, and added to the set of labelled samples $\mathscr{X}_l$. The detection model $\Theta$ is then trained for a fixed number of epochs using the expanded set, $\mathscr{X}_l$. The reward for the policy network for selecting the samples is the performance of the underlying model $\Theta$ computed using a metric apropos to the task (e.g. \textit{Average Precision} for layout detection, and \textit{F-score} for named entity recognition) on a separate held-out set, $\mathscr{X}_{met}$. The training of the policy $\Pi$ is performed through episodes of active learning. 

\begin{table}[!h]
\centering
\begin{tabular}{p{0.3\linewidth} | p{0.6\linewidth}}
\hline
\textbf{Notations} & \textbf{Description} \\ \hline \hline
$\mathscr{X}_{train}$, $\mathscr{X}_{val}$, $\mathscr{X}_{test}$  & Train, Validation and Test sets of a given dataset\\ \hline
$\mathscr{X}_u$, $\mathscr{X}_l$, $\mathscr{X}_{init}$ & Unlabelled, labelled, and initial labelled sets \\ \hline
$\mathscr{X}_{cand}$ & Candidate unlabelled examples for an active learning cycle\\ \hline
$\mathscr{X}_{met}$, $\mathscr{X}_{state}$ &  Metric calculation set, State representation set\\ \hline
$\mathcal{A}_t$, $\mathcal{S}_t$, $\mathcal{R}_t$ & Action, State and Reward at time $t$\\ \hline
$\Pi$, $\Theta$ & Policy deep Q network and Prediction model to be trained \\ \hline
$\mathbb{M}$, $\mathbb{B}$ & Memory buffer for Q learning, Total budget for active learning \\ \hline
$n_{cycle}$, $n_{pool}$, $n_{init}$ & Number of samples to be acquired in one active learning cycle, Number of samples in a pool, Number of samples labelled for initial training \\ \hline
\end{tabular}
\caption{Notations used to represent various data splits and model components.}
\label{tab:notations}
\end{table}

We now describe various components of the proposed policy-based active learning approach in details. 
\subsection{Data Splits}
Given a dataset $\mathbb{D}$, we split the samples (or use the existing splits of the dataset) into $\mathscr{X}_{train}$, $\mathscr{X}_{val}$, and $\mathscr{X}_{test}$ sets. For the two stages of \textit{OPAD}, the further splits are as follows. 
\paragraph{\textbf{During policy training stage}}
We separate a set of samples $\mathscr{X}_{met}$ along with their labels from $\mathscr{X}_{train}$, which is used for validating the performance of underlying model $\Theta$ and computing rewards for training the policy DQN $\Pi$. For the RL setup of the policy DQN, we use a held-out set $\mathscr{X}_{state}$ which is used together with $\mathscr{X}_{cand}$ later to compute overall state representation. Note that, unlike \cite{casanova2020Reinforced}, we do not require labels for $\mathscr{X}_{state}$, which further reduces the annotation budget. During this stage, we train the detection model $\Theta$ on $\mathscr{X}_{l}$, which is initialized with $\mathscr{X}_{init}$ and populated with samples from $\mathscr{X}_{u}$ as the active learning progresses. Here, $\mathscr{X}_{init}$ is a set with $n_{init}$ randomly selected samples with the corresponding labels for initial training of the model $\Theta$. Therefore, before the active learning process starts, $\mathscr{X}_{u}$ equals $\mathscr{X}_{train} - \{\mathscr{X}_{init} + \mathscr{X}_{state} + \mathscr{X}_{met}\}$, and $\mathscr{X}_{l}$ equals $\mathscr{X}_{init}$. 

\paragraph{\textbf{During deployment stage}}
We utilize the $\mathscr{X}_{val}$ set for training the detection model $\Theta$. We make this differentiation from the policy training stage to ensure that sample selection by the policy happens on an unseen set. During this stage, we use the same terminology $\mathscr{X}_{init}$, $\mathscr{X}_{l}$, and $\mathscr{X}_{u}$ from the previous stage. However, the $n_{init}$ samples in $\mathscr{X}_{init}$ set are selected from the $\mathscr{X}_{val}$ set and therefore, at the start of the active learning process $\mathscr{X}_{u}$ equals $\mathscr{X}_{val} - \{\mathscr{X}_{init}\}$, and $\mathscr{X}_{l}$ equals $\mathscr{X}_{init}$.  We use the same set of examples for the state computation set $\mathscr{X}_{state}$. In this stage we do not require the $\mathscr{X}_{met}$ set. \newline

\noindent Though we have ground truth annotations available for all the samples in all the three sets, to simulate the annotation setup, we mask this data from both $\Theta$ and $\Pi$ models and utilize the labels as and when required.
\subsection{Active Learning}
\begin{figure*}
\centering
  \includegraphics[width=0.85\textwidth]{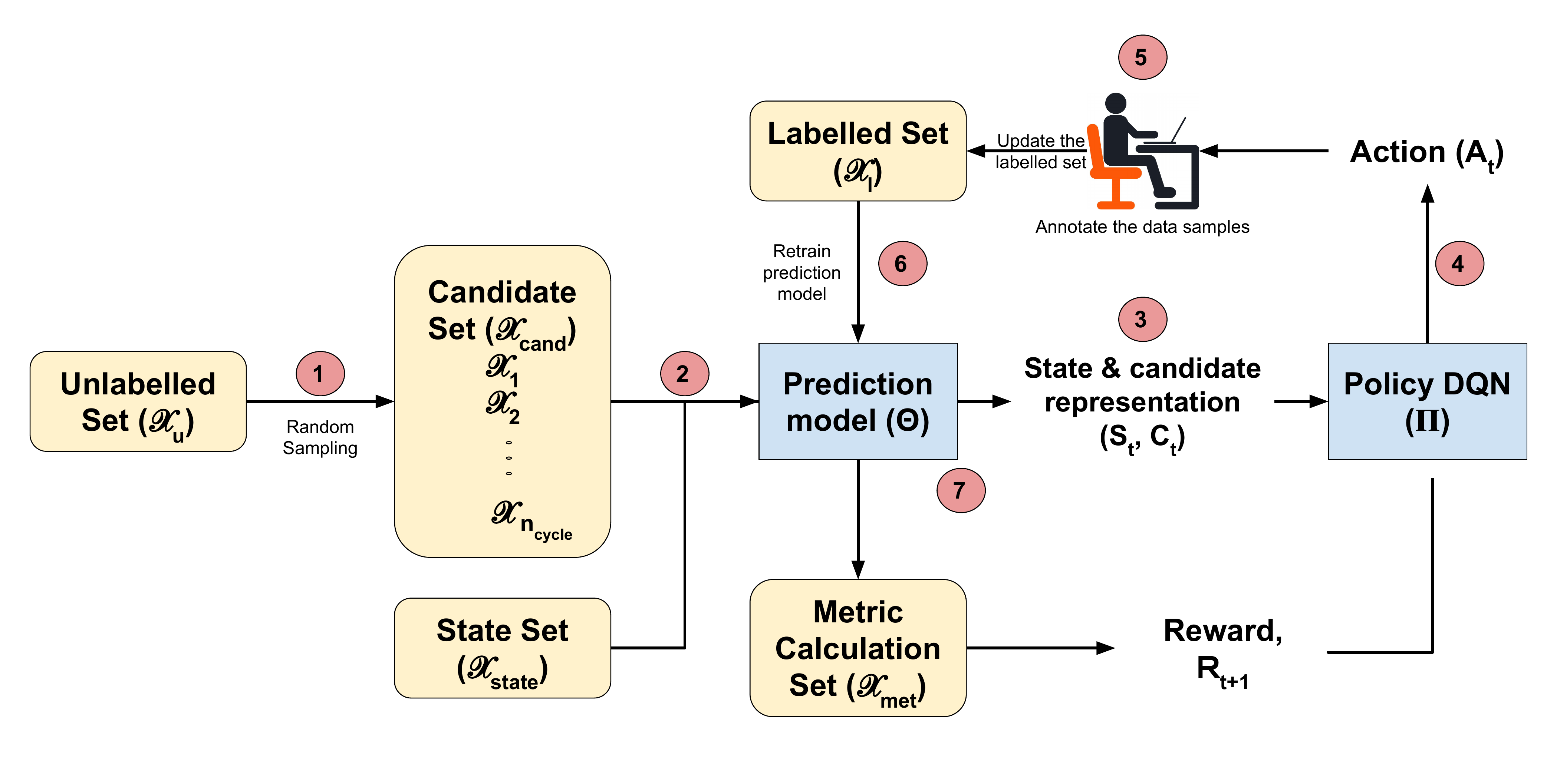}
  \caption{\textbf{Overview of the policy training in \textit{OPAD}} - (1) Candidate samples are chosen randomly from the unlabelled pool $\mathscr{X}_u$. (2) State representation is calculated using $\mathscr{X}_{cand}$ and $\mathscr{X}_{state}$, which is then passed to the policy DQN $\Pi$ to select the samples to be annotated (3, 4 and 5). (6) The labelled set $\mathscr{X}_l$ is then updated and the model $\Theta$ is retrained. (7) Finally, reward is computed using the set $\mathscr{X}_{met}$.}
\label{fig:modeloverview}
\end{figure*}
\begin{figure}[!h]
\centering
    \includegraphics[width=1.0\linewidth]{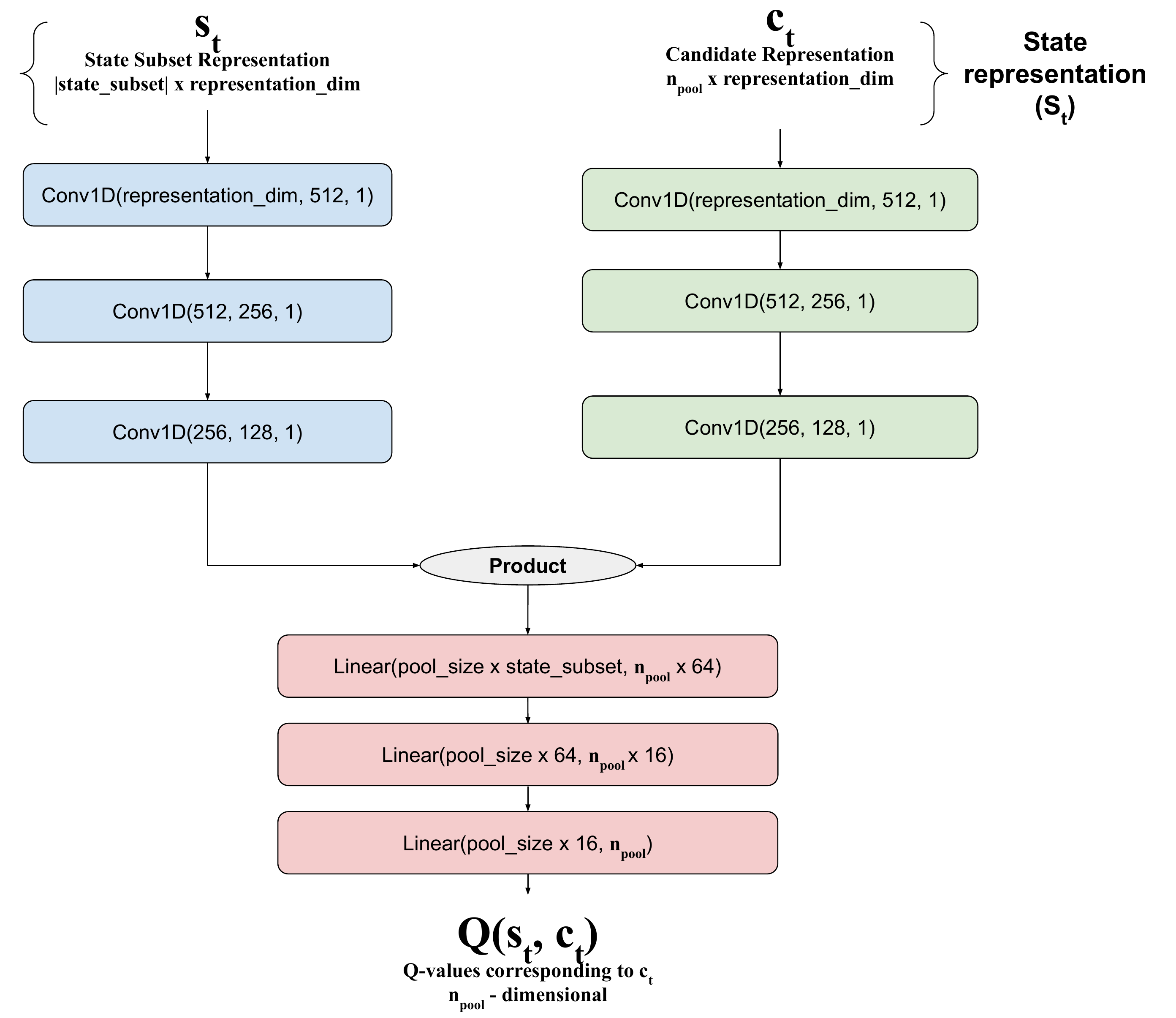}
   \caption{Architecture of the proposed Deep Query Network, $\Pi$ for the policy.}
\label{fig:dqn}
\end{figure}
\begin{algorithm}[!h]
\begin{algorithmic}[1]
\caption{Training the Policy DQN, $\Pi$}
  \Statex \noindent\textbf{Input:} $\mathscr{X}_{train}$, budget $\mathbb{B}$
  \Statex \textbf{Output:} Policy DQN, $\Pi$, trained for querying the samples for annotation
  \State Randomly sample examples from $\mathscr{X}_{train}$ to form $\mathscr{X}_{state}$ and $\mathscr{X}_{met}$ sets.
  \State Initialize policy and target DQN
  \State Initialize memory replay buffer $\mathbb{M}$
  \While{convergence of DQN loss}
    \State Initialize $\Theta$
    \State Randomly sample $n_{init}$ from $\mathscr{X}_{train} - \{\mathscr{X}_{state} + \mathscr{X}_{met}\}$ to form $\mathscr{X}_{init}$
    \State Initialize $\mathscr{X}_{u}$ to $\mathscr{X}_{train} - \{\mathscr{X}_{state} + \mathscr{X}_{met} + \mathscr{X}_{init}\}$
    \State Initialize $\mathscr{X}_{l}$ to $\mathscr{X}_{init}$
    \State Train the model $\Theta$ on $\mathscr{X}_l$
    \State Compute the performance metric on $\mathscr{X}_{met}$
    \While{Consumption of budget $\mathbb{B}$}
    \State Sample $n_{pool} \times n_{cycle}$ number of samples from $\mathscr{X}_u$ as candidates for labelling $\mathscr{X}_{cand}$
    \State Compute state representation $\mathcal{S}_t$ using predictions of model $\Theta$ on $\mathscr{X}_{state}$ and $\mathscr{X}_{cand}$
    \State Select $n_{cycle}$ samples from $\mathscr{X}_{cand}$ using $\epsilon$-greedy policy and add it to $\mathscr{X}_{l}$ - Action $\mathcal{A}_{t}$
    \State Retrain the model $\Theta$ on $\mathscr{X}_{l}$
    \State Compute the metric on the $\mathscr{X}_{met}$
    \State Compute the reward $\mathcal{R}_{t+1}$ as the difference in metric
    \State Re-do steps 14 and 15 - Next State $\mathcal{S}_{t+1}$
    \State Add tuple ($\mathcal{S}_{t}$, $\mathcal{A}_{t}$, $\mathcal{R}_{+1}$, $\mathcal{S}_{t+1}$) to the memory replay buffer $\mathbb{M}$
    \State Optimize policy DQN, $\Pi$
    \EndWhile
  \EndWhile
\label{algo:pbal}
\end{algorithmic}
\end{algorithm}
Figure \ref{fig:modeloverview} shows an overview of active learning (\textit{inner while loop} at step 11 in Algorithm \ref{algo:pbal}) in a single episode of policy training. In an active learning cycle, we select $n_{pool}\times n_{cycle}$ number of samples from the set $\mathscr{X}_u$, which represent the candidates selected for the current active learning cycle $\mathscr{X}_{cand}$. 
The policy DQN $\Pi$ computes $Q$-value for samples within each pool containing $n_{pool}$ samples, based on candidate set $\mathscr{X}_{cand}$ and state representation set $\mathscr{X}_{state}$. The policy selection network is optimized to maximize the reward, $\mathcal{R}_t$:
\begin{equation}
\label{eq:policy_selection}
    Q^*(\mathcal{S}_{t},\mathcal{A}_{t}) = \max_{\Pi} \mathbb{E}[\mathcal{R}_{t+1} | \mathcal{S}_{t},\mathcal{A}_{t}, \Pi]
\end{equation}
The annotator then annotates the selected samples, and the labelled set $\mathscr{X}_l$ is updated by adding these new samples. We then retrain the model $\Theta$ using the updated labelled set and finally calculate the reward for the current cycle $\mathcal{R}_t$ by measuring the performance of the model $\Theta$ on $\mathscr{X}_{met}$.
\begin{equation}
\label{eq:vanilla_reward}
    \mathcal{R}_{t+1} = Performance_{t, \mathscr{X}_{met}} - Performance_{t-1, \mathscr{X}_{met}}
\end{equation}
where $Performance$ is measured in terms of \textit{AP metric} for layout and object detection tasks, and \textit{F-score} for named entity recognition task. Algorithm \ref{algo:pbal} summarizes the training phase of the proposed approach. 

\subsection{Policy Training Stage}
\paragraph{\textbf{Policy Network}}
Our policy network $\Pi$ is a deep query network, as shown in Figure \ref{fig:dqn}. 
The underlying prediction model $\Theta$ computes the representations $\mathbf{c_t}$ and $\mathbf{s_t}$ from the sets $\mathscr{X}_{cand}$ and $\mathscr{X}_{state}$ respectively (details in Section \ref{sec:mdp_rep}). The policy network then receives the two inputs $s_t$, and $c_t$, which we denote as the state representation $\mathcal{S}_t$ in Figure \ref{fig:dqn}. We pass the two representations through convolution layers, followed by vector product of state and candidate representations. The final Q-value is obtained by passing the combined representation through fully connected layers.

\paragraph{\textbf{Policy Optimization}}
The computed Q-value is used for selecting $n_{cycle}$ samples at each step. For this, a memory or experience replay buffer, $\mathbb{M}$ is created using MDP state representation tuples, ($\mathcal{S}_{t}$, $\mathcal{A}_{t}$, $\mathcal{R}_{t+1}$, $\mathcal{S}_{t+1}$). Further, as a batch of $n_{cycle}$ needs to be selected, the candidate set, $\mathcal{X}_{cand}$, is randomly partitioned into $n_{cycle}$ mini-batches, and action set $\mathcal{A}_t$ is set to ${\mathcal{A}^i_t}^{n_{cycle}}_{i=1}$. The loss is then optimized as follows to train the policy network:
\begin{equation}
\label{eq:policy_network_loss}
    Loss(\Pi) = \mathbb{E}_{t \in \mathbb{M}}[(\mathcal{Y}^i_t - Q(\mathcal{S}_t, \mathcal{A}^i_t)); \Pi))^2]
\end{equation}
The values for $\mathcal{Y}^i_t$ are computed using a double DQN formulation~\cite{haussmann2019CLassRLAL} incorporating a target network, $\Pi'$ for stable training:
\begin{equation}
\label{eq:double_dqn_policy}
\mathcal{Y}^i_t = \mathcal{R}_{t+1} + \max_{\mathcal{A}^i_{t+1}}\gamma Q(\mathcal{S}_{t+1}, \mathcal{A}^i_{t+1}; \Pi' ); \Pi) 
\end{equation}
where, $\gamma$ is the discount factor for future reward, set to $0.9$ in our experiments.

\paragraph{\textbf{$\epsilon$-greedy selection}}
To encourage exploration of diverse samples by the policy during training, an $\epsilon$-greedy strategy is followed while training the policy, which selects a random sample for the action $\mathcal{A}^i_t$ with probability $epsilon$, instead of the sample maximizing Q-value. The $\epsilon$ value starts with $0.9$ for the initial cycle, and decreases by a factor of $0.1$ for subsequent cycles. For policy deployment, $\epsilon$ is set to 0. The gradient optimization is done using the temporal difference method~\cite{sutton1988learning}.

\subsection{Deployment Stage}
\begin{algorithm}
\begin{algorithmic}[1]
\caption{Testing the Policy DQN, $\Pi$}
  \Statex \noindent\textbf{Input:} $\mathscr{X}_{val}$, $\mathscr{X}_{test}$, $\mathscr{X}_{state}$, budget $\mathbb{B}$
  \State Randomly sample $n_{init}$ from $\mathscr{X}_{val}$ to form $\mathscr{X}_{init}$
  \State Initialize $\mathscr{X}_{u}$ to $\mathscr{X}_{val} - \{\mathscr{X}_{init}\}$
  \State Initialize $\mathscr{X}_{l}$ to $\mathscr{X}_{init}$
  \State Initialize $\Theta$
  \State Train the model $\Theta$ on $\mathscr{X}_l$
  \State Compute the performance metric on $\mathscr{X}_{test}$
  \While{Consumption of budget $\mathbb{B}$}
    \State Sample $n_{pool} \times n_{cycle}$ number of samples from $\mathscr{X}_u$ as candidates for labelling $\mathscr{X}_{cand}$
    \State Compute state representation $\mathcal{S}_t$ using predictions of model $\Theta$ on $\mathscr{X}_{state}$ and $\mathscr{X}_{cand}$
    \State Select $n_{cycle}$ samples from $\mathscr{X}_{cand}$ using $\epsilon$-greedy policy and add it to $\mathscr{X}_{l}$ - Action $\mathcal{A}_{t}$
    \State Retrain the model $\Theta$ on $\mathscr{X}_{l}$
    \State Compute the metric on the $\mathscr{X}_{test}$ and report
  \EndWhile
\label{algo:pbal_test}
\end{algorithmic}
\end{algorithm}
Algorithm \ref{algo:pbal_test} summarizes the deployment stage (or policy testing stage). We freeze the parameters of the model $\Pi$ in this stage. We use the $\mathscr{X}_{val}$ set to iteratively select the samples and train the model $\Theta$. At the end of each active learning cycle we compute the performance of the model $\Theta$ on the held-out set $\mathscr{X}_{test}$ and report the values in Section \ref{sec:experiments}. 
\subsection{Weak labelling}
\label{sec:weak_labeling}
\begin{figure}[!h]
\centering
    \includegraphics[width=1.0\linewidth]{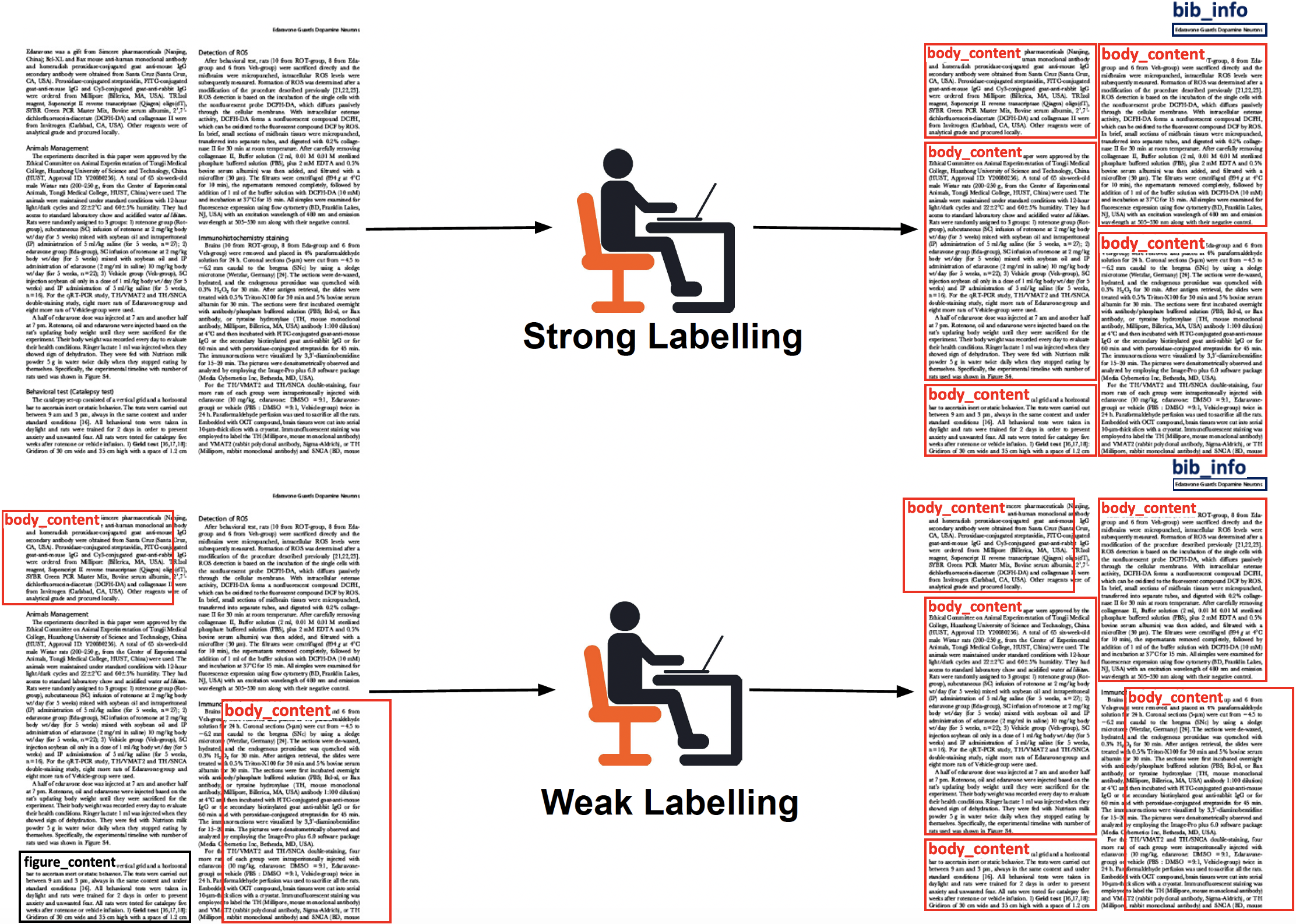}
   \caption{\textbf{Weak labelling in the case of layout detection.} In the top image, the annotator has to draw and mark all the layout boxes, while in the bottom image, the annotator can verify the predictions of the model in the input image, and add new boxes. Image is best viewed in color.}
\label{fig:weaklabeling}
\end{figure}

In a usual annotation scenario (as shown in Figure \ref{fig:weaklabeling} - top), the annotator has to mark all the entities present in a sample by drawing the bounding boxes and selecting labels for them. To reduce the annotation cost, we propose a weak labelling annotation framework (Figure \ref{fig:weaklabeling} - bottom). Inspired from \cite{papadopoulos2016NoBBoxesHumanVerification}, the annotator is shown the document as well as the predictions with high confidence from the model $\Theta$ for that document. The annotator can then (1) add a missing box, (2) mark a box either correct or incorrect, and (3) mark a label either correct or incorrect for the associated box. The annotation interface for the weak labelling approach is shown in Figure \ref{fig:annotationinterface}. 

The advantage of weak labelling is that it significantly reduces the annotation time. Annotation of a new entity by drawing a bounding box or selecting words takes $\sim 15$ seconds on an average in the case of detection tasks and $\sim 4$ seconds in case of named entity recognition. Verifying an entity takes $\sim 5$ seconds for layout detection task and $\sim 2$ seconds for named entity recognition\footnote{All the mentioned values are average annotation times of $3$ individuals measured on the developed annotation tool}.

\subsection{Additional Rewards} We propose the following additional rewards to improve the performance of the active learning approach.
\begin{itemize}
    \item \textbf{Class balance reward:} To reduce class imbalance in the newly acquired samples that are to be labelled, $\mathscr{X}_{new}$, we propose an additional class distribution entropy reward which reinforces a class-balanced selection of samples.
        \begin{equation}
        \label{eq:class_ent_reward}
            \mathcal{R}_{cls\_ent} = \mathcal{H}(P(\mathscr{X}_{new}))
        \end{equation}
        where $\mathcal{H}$ is the Shannon entropy function~\cite{shannon2001mathematical}, and $P(\mathscr{X}_{new})$ is the probability distribution over various classes for the newly acquired samples $\mathscr{X}_{new}$.
    \item \textbf{Human feedback reward:} In a weak labelling scenario, where the annotator can modify the output from the prediction model, $\Theta$, a human feedback signal could be added at each active learning cycle while training the policy. The objective is to promote the selection of those samples for which the annotator modifies the high confidence predictions of $\Theta$ heavily because such samples would be more informative for the model $\Theta$. Accordingly, the additional human feedback reward for detection during training time is given as,
        \begin{equation}
        \label{eq:feedback_reward}
            \mathcal{R}_{feedback} = AP_{after\_feedback} - AP_{before\_feedback}
        \end{equation}
        where $AP_{after\_feedback}$ is the AP metric on the newly acquired samples, after the annotator has verified the predictions, and $AP_{before\_feedback}$ is the AP of the samples before feedback. 
\end{itemize}

\section{Experiments and Results}
\label{sec:experiments}
In this section, we provide a comprehensive experimental evaluation of the proposed policy-based active learning approach on the document understanding tasks, document layout detection and named entity recognition. Furthermore, we also evaluate our models on Pascal VOC object detection task to demonstrate the generalizability of the proposed solution across different domains.

\subsection{Datasets}
We use the following datasets for the corresponding tasks:
\begin{itemize}
    \item \textbf{GROTOAP2}~\cite{Tkaczyk2014GROTOAP2T} dataset is used for the complex document layout detection task. The dataset consists of 22 layout classes for scientific journals. We sampled two sets of $5000$ images as training and validation sets. Among these, we hold-out 10\% for reward computation set $\mathscr{X}_{met}$ and 256 random samples for $\mathscr{X}_{state}$ and use the remaining samples for the active learning setup. We use the validation set for simulating the active learning during the deployment phase and finally report the performance on a held-out subset of $2500$ images. Further, we merged those classes having very few instances (e.g. \textit{glossary}, \textit{equation}, etc.) with the \textit{body content} class, resulting into a modified dataset with 13 classes.
    \item \textbf{CoNLL-2003}~\cite{tjong2003introduction} English Corpus is used for performing active learning experiments for the named entity recognition task. We use the $14042$ sentences from the \textit{train} set of CoNLL-2003 training our policy in active learning setup after separating out 10\% for $\mathscr{X}_{met}$ and 512 sentences for $\mathscr{X}_{state}$. We use the $3251$ sentences of the dev set of this corpus to train the underlying $\Theta$ through active learning cycles in the deployment stage. Finally, we use the \textit{test} set of CoNLL-2003 consisting of $3454$ sentences for calculating and reporting the F-scores of the $\Theta$ model during the deployment stage.
    \item \textbf{Pascal VOC-2007}~\cite{pascal-voc-2007} dataset with 20 object classes is used for the object detection task. We use the \textit{train} set of VOC-2007 containing $2501$ images during the policy training phase. Similar to layout detection task, we hold-out 10\% for reward computation set $\mathscr{X}_{met}$ and 256 random samples for $\mathscr{X}_{state}$ and use remaining samples for the active learning setup. During the deployment phase, we utilize the \textit{val} set of VOC-2007 containing $2510$ images for simulating the active learning setup i.e selecting samples using trained $\Pi$ model and training the model $\Theta$. We use the \textit{test} set of VOC-2007 consisting of --- samples for reporting the performance of model $\Theta$ after each active learning cycle during the deployment stage.
\end{itemize}

\noindent We also use the following datasets for pre-training the underlying model $\Theta$:
\begin{itemize}
    \item \textbf{PubLayNet}~\cite{zhong2019publaynet} We use this dataset for pre-training $\Theta$ for document layout detection. This dataset contains over 360K page samples and has typical document layout elements such as  \textit{text}, \textit{title}, \textit{list}, \textit{figure}, and \textit{table} as the annotations. While the \textit{list}, \textit{figure}, \textit{table} and \textit{title} classes contains the corresponding information from document, the \textit{text} category consists of the rest of the content such as author, author affiliation; paper information; copyright information; abstract; paragraph in main text, footnote, and appendix; figure \& table caption; table footnote. 
    \item \textbf{MS-COCO}~\cite{lin2014microsoft} This dataset consists of 91 object classes. We use this dataset to pre-train the underlying classification model $\Theta$ (i.e. Faster-RCNN model) in the case of object detection on the VOC dataset. We pre-train the model $\Theta$ on this dataset and remove the last layers from both the class prediction and bounding box regression branches which are class-specific.  
    
\end{itemize}

\subsection{Models and configurations}
We use the Faster-RCNN model~\cite{ren2015faster} with RESNET-101 backbone\footnote{https://github.com/facebookresearch/detectron2}~\cite{he2015Deep} as the underlying prediction model for the layout detection and object detection tasks. The Faster-RCNN model is pre-trained on a subset of $15000$ images from PubLayNet~\cite{zhong2019publaynet} dataset for the layout detection task, and on MS-COCO~\cite{lin2014microsoft} dataset for the object detection task to bootstrap the active learning experiments. For the NER task, we use the BiLSTM-CRF~\cite{huang2015bidirectional} model for recognition task. 

For active learning, we use a seed set of $512$ labelled samples in case of detection tasks and $100$ labelled samples in case of NER for training the prediction model initially. Both the Faster-RCNN model and BiLSTM-CRF models are trained for $1000$ iterations on the labelled set in an active learning cycle. In each of the $10$ active learning cycles we select $64$ samples in the case of detection tasks and $25$ samples in the case of NER, from unlabelled dataset for labelling giving a total of $1152$ and $350$ labelled samples in a single episode for detection tasks and NER respectively. We run $10$ episodes of these active learning cycles to train the policy network. The learning rate for training the policy DQN is set to $0.001$ with a gamma value of $0.998$. The learning rates of Faster-RCNN and BiLSTM-CRF are set to $0.00025$ and $0.01$ respectively. We also apply a momentum of $0.95$ to optimize the training of policy network. We set the size of memory replay buffer $\mathbb{M}$ to 1000 samples with first-in-first-out mechanism. 

\begin{figure*}[!th]
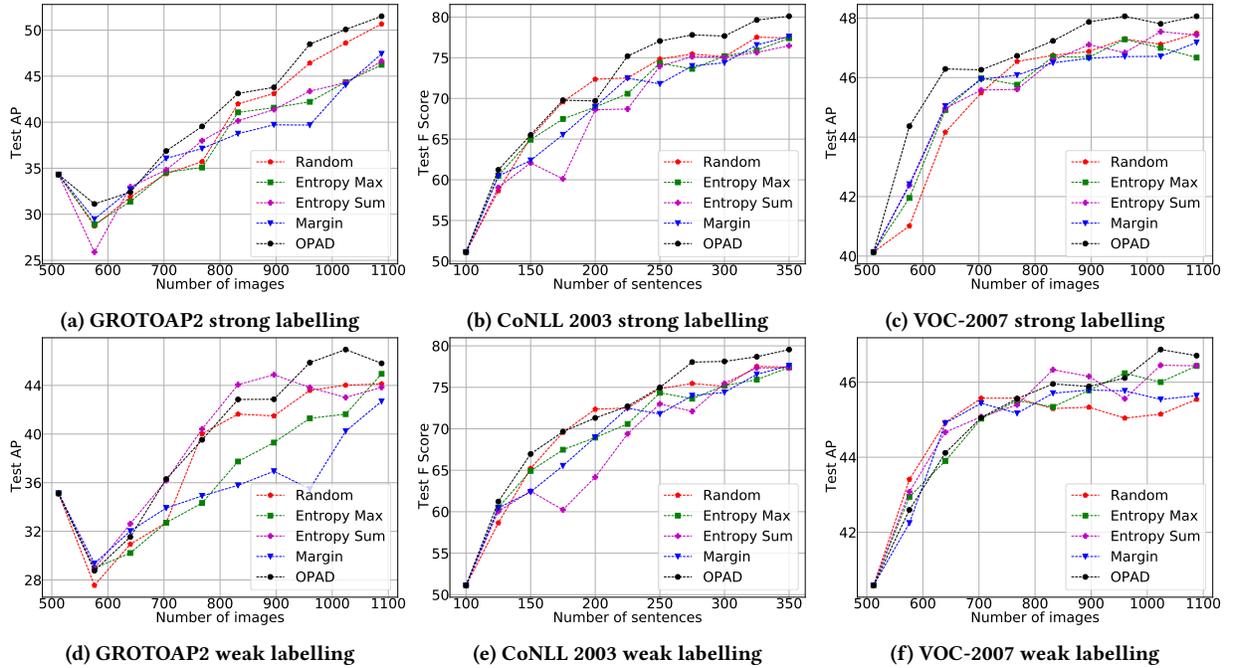

    \centering
    \begin{subfigure}[b]{0.30\linewidth}
        \includegraphics[scale=0.03]{imgs/plots/groto_strong.pdf}
        \caption{GROTOAP2 strong labelling}
        \label{fig:groto_strong}
    \end{subfigure}
    ~
    \begin{subfigure}[b]{0.30\linewidth}
        \includegraphics[scale=0.03]{imgs/plots/ner_strong.pdf}
        \caption{CoNLL 2003 strong labelling}
        \label{fig:ner_strong}
    \end{subfigure}
    ~
    \begin{subfigure}[b]{0.30\linewidth}
        \includegraphics[scale=0.03]{imgs/plots/voc_strong.pdf}
        \caption{VOC-2007 strong labelling}
        \label{fig:voc_strong}
    \end{subfigure}
    \\
    \begin{subfigure}[b]{0.30\linewidth}
        \includegraphics[scale=0.03]{imgs/plots/groto_weak.pdf}
        \caption{GROTOAP2 weak labelling}
        \label{fig:groto_weak}
    \end{subfigure}
    ~
    \begin{subfigure}[b]{0.30\linewidth}
        \includegraphics[scale=0.03]{imgs/plots/ner_weak.pdf}
        \caption{CoNLL 2003 weak labelling}
        \label{fig:ner_weak}
    \end{subfigure}
    ~
    \begin{subfigure}[b]{0.30\linewidth}
        \includegraphics[scale=0.03]{imgs/plots/voc_weak.pdf}
        \caption{VOC-2007 weak labelling}
        \label{fig:voc_weak}
    \end{subfigure}
    
    \caption{Plots showing the performance of the methods, \textit{viz.} random, entropy, margin and proposed, for GROTOAP2, CoNLL-2003 and VOC-2007 for both strong and weak labelling settings.}
    \label{fig:graph_plots}
\end{figure*}

\subsection{MDP state representation} 
\label{sec:mdp_rep}
For the layout detection and object detection tasks, we use a randomly sampled set of $256$ images from the \textit{train} set as the subset for representing the overall distribution of the dataset ($\mathscr{X}_{state}$). We pass each instance from the candidate ($\mathscr{X}_{cand}$) and state ($\mathscr{X}_{state}$) subsets through the Faster-RCNN model, to get the top $50$ confident bounding box predictions. We concatenate the class scores for these top $50$ predictions to the feature map of RESNET-101 backbone to get a final representation ($1256$-dimension for VOC-2007, and $906$-dimension for GROTOAP2) for each sample in the candidate and state subset sets.

For NER task, we use a randomly sampled set of $512$ sentences from the \textit{train} set of CoNLL-2003 as the $\mathscr{X}_{state}$ set. We pass each sentence from the $\mathscr{X}_{cand}$ and $\mathscr{X}_{state}$ sets through the BiLSTM-CRF model and compute the final representation by taking the class scores of all the entities in the sentence. We pad each sentence to get a $150$ dimensional representation, and with $20$ classes using IOBES formatting~\cite{ramshaw1999text}, we generate a $3000$-dimensional representation for each sentence. The representations thus obtained from the samples in $\mathscr{X}_{cand}$ are stacked to form $\mathbf{c_t}$, and similarly $\mathbf{s_t}$ from the set $\mathscr{X}_{state}$. Together $\mathbf{c_t}$ and $\mathbf{s_t}$ form the state representation $\mathcal{S}_t$ in Figure \ref{fig:dqn}.

\subsection{Human Annotation Simulation}
To simulate the role of a human annotator for weak labelling, we use the ground truths of the datasets on which we perform our experiments. In detection tasks (i.e. layout detection and object detection), we consider the predictions which have an IoU greater than 0.5 with the ground truth box as the boxes being marked as correct by the annotator. For those boxes in the ground truth which do not have any prediction with IoU greater than 0.5, we include that box into the labelled set marking as a full annotation (a strong label). In case of named entity recognition, we compare the predictions with the ground truth, and those predictions which are correct are included in the weakly labelled set. The ground truth entities which are missed are added as a strong label.
\begin{table}
\centering
\scalebox{0.90}{
    \begin{tabular}{c|c|c|c|c}
    \toprule
        & & \multicolumn{3}{c}{Avg time (seconds)$\longrightarrow$} \\
        \cline{3-5}
        & & & \\
        & Method$\downarrow$ & GROTOAP2 & CoNLL2003 & VOC2007 \\
    \midrule
        \multirow{4}{*}{\rotatebox{90}{{Strong}}} & Random & 10m14s & 0m29s & 12m21s \\
        & Entropy Max & 18m14s & 1m15s & 17m16s \\
        & Entropy Sum & 18m01s & 1m10s & 15m53s \\
        & Margin & 18m00s & 1m05s & 15m39s \\
        & OPAD & 11m22s & 0m53s & 14m00s \\
    \midrule
        \multirow{4}{*}{\rotatebox{90}{{Weak}}} & Random & 10m23s & 0m35s & 12m24s \\
        & Entropy Max & 18m20s & 1m20s & 17m31s \\
        & Entropy Sum & 18m10s & 1m11s & 15m26s \\
        & Margin & 18m03s & 1m10s & 15m48s \\
        & OPAD & 11m36s & 1m00s & {{14m12s}} \\
    \bottomrule
    \end{tabular}
    }
    \caption{Time required to complete one active learning cycle i.e selection of samples for various algorithms along with the training time of the model $\Theta$. Note that the training time of the model $\Theta$ is constant across the algorithms and hence the relative order is representative of the sample selection time. OPAD performs the best next to the simple random algorithm.}
     \label{tab:annot_time}
\end{table}

\begin{table}[!h]
\centering
\scalebox{0.95}{
    \begin{tabular}{c|c|c|c|c}
    \toprule
        & & \multicolumn{3}{c}{Annotation time required(seconds)$\longrightarrow$} \\
        \cline{3-5}
        & & & \\
        & Method$\downarrow$ & GROTOAP2 & CoNLL2003 & VOC2007 \\
    \midrule
        \multirow{4}{*}{\rotatebox{90}{{Strong}}} & Random & 72500 & 1650 & 9000 \\
        & Entropy Max & 81600 & 2100 & 10500 \\
        & Entropy Sum & 81200 & 2050 & 10000 \\ 
        & Margin & 92700 & 1950 & 11000 \\
        & Ours & { \textbf{66000}} & { \textbf{1000}} & { \textbf{7000}} \\
    \midrule
        \multirow{4}{*}{\rotatebox{90}{{Weak}}} & Random & 38000 & 1100 & 4250 \\
        & Entropy Max & 39000 & 1500 & {\textbf{2000}} \\
        & Entropy Sum & 41000 & 1600 & 2500 \\ 
        & Margin & 48000 & 1300 & 7500 \\
        & Ours & { \textbf{33000}} & { \textbf{700}} & {{2250}} \\
    \bottomrule
    \end{tabular}
    }
    \caption{Annotation time required to reach an AP of 42.5 on GROTOAP2, an F1 score of 76.0 on CoNLL2003, and an AP of 45.5 on VOC-2007. These values indicate the minimum achievable best performances by all the models on the datasets.}
     \label{tab:time_for_perform}
\end{table}

\subsection{Results}
We compare the performance of our proposed method with three baselines - 
\begin{itemize}
    \item \textbf{Random} Data samples from the unlabelled pool are randomly chosen for annotation. 
    \item \textbf{Entropy}~\cite{roy2018DetectionHeuristicAL} For the entropy-based selection, first the entropy of class prediction probability by $\Theta$ is computed over all the entities of a data sample. We present results for aggregating entropy of a single sample in two ways: 1. maximum entropy, 2. sum of entropy of all detected entities within the sample, and then the samples with the highest aggregate entropy are selected for labelling.
    \item \textbf{Margin}~\cite{brust2019DetectionHeuristicAL} Similar to entropy, a $v_{1vs2}$ margin score is computed using the difference of prediction probability of highest and second highest class for all the instances of a sample. Then, the maximum margin score over all the instances is taken to be the aggregate margin measure for the sample. Samples with the highest aggregate margin are selected for labelling. The baseline metrics are as described in the existing prior art.  
\end{itemize}

Though there are active learning approaches in the literature that have been proposed for content detection tasks, they impose implicit or explicit constraint of detecting only one object/element or a fixed number of objects/elements per input instance to generate a viable fixed length representations of instances for their active learning setups. However, in our proposed active learning setup, we impose no such restrictions on the model and utilize all the detections from the underlying model $\Theta$, making those baseline models infeasible for a direct comparison. Figure~\ref{fig:graph_plots} shows the accuracy of all the methods on the test sets of different datasets, for both strong and weak labelling settings. We can observe that the proposed policy-based AL method significantly outperforms the baseline methods. This is because of the optimized selection policy, learned to reward the better performance of the prediction model. While the curves for VOC-2007 and CoNLL-2003 datasets approach saturation, we stop the GROTOAP2 training before reaching saturation as our objective is to show the performance of the underlying model with a limited budget. Note that the proposed method uses vanilla reward in all the plots in Figure \ref{fig:graph_plots}. Further, as shown in Table \ref{tab:annot_time} and Table \ref{tab:time_for_perform}, the proposed method takes significantly less time for annotation than the baselines to reach the minimum best performance achievable by all the models, while performing only next to random algorithm for sample selection timings. Please note that the annotation times in Table \ref{tab:time_for_perform} are based on the number of samples selected for annotation multiplied by the average human annotation times mentioned in Section \ref{sec:weak_labeling}.
\section{Ablation Study}
In this section we discuss the importance of the proposed additional rewards in improving the performance of the proposed AL approach. 
\subsection{Class balance reward}
We conduct ablations by adding the class distribution entropy reward (Equation \ref{eq:class_ent_reward}) to the vanilla reward function. The overall reward function is:
\begin{equation}
\label{eq:cls_ent_combined}
    \mathcal{R}_{overall} = \mathcal{R}_{t} + \lambda * \mathcal{R}_{cls\_ent}
\end{equation}
where $\lambda$ is a hyper-parameter, and $\mathcal{R}_{t}$ is the vanilla reward. We compare the results of using this reward in our policy against the baselines and vanilla policy in a strong labelling setting in Table \ref{tab:clsent}. We can observe a significant increase in performance, specifically on NER task, with the overall reward as compared to the vanilla reward policy. 
This experiment clearly shows that there is a substantial need to remove the class imbalance in the samples selected through the policy.

\begin{table}[!h]
\centering
\scalebox{0.95}{
\begin{tabular}{lccc}
\toprule
 & \multicolumn{2}{c}{AP} & F-score\\ 
{Method $\downarrow$}         & GROTOAP2 & CoNLL-2003 & VOC-2007 \\
\midrule
Random                  & 50.668 & 77.438 & 47.490 \\
Entropy Max             & 46.229 & 77.401 & 46.671 \\
Entropy Sum             & 46.634 & 77.351 & 47.431 \\
Margin                  & 47.428 & 77.598 & 47.179 \\
OPAD                    & 51.508 & 80.099 & 48.061 \\
OPAD (ClsEnt $\lambda$ - 0.25) & \textbf{53.241} & \textbf{86.853} & 47.727  \\
OPAD (ClsEnt $\lambda$ - 0.50) & 51.185 & 86.541 & 47.701 \\
OPAD (ClsEnt $\lambda$ - 0.75) & 52.143 & 86.512 & \textbf{48.566} \\
OPAD (ClsEnt $\lambda$ - 1.0)  & 51.530 & 86.395 & 48.060 \\ 
\bottomrule
\end{tabular}}
\caption{Performance of our method on test data with class distribution entropy reward on various datasets. The reported results are after consuming a total budget of 1152 samples for GROTOAP2 and VOC-2007, and 350 samples for the CoNLL-2003 datasets.}
\label{tab:clsent}
\end{table}


\subsection{Human feedback reward}
In this experiment we report the effect of adding human feedback to the vanilla reward, i.e.
\begin{equation}
\label{eq:human_feedback_combined}
    \mathcal{R}_{overall} = \mathcal{R}_{t} + \lambda * \mathcal{R}_{feedback}
\end{equation}
where $\lambda$ is a hyper-parameter. We report the results of using this overall reward in our policy in Table \ref{tab:feedback}, along with the baselines and vanilla policy in a weak labelling setup. We observe that having a small weight on the feedback reward results in a jump in the performance on VOC-2007. Thus, the reward can be useful in some cases; however, further investigation is needed to establish its efficacy. Note that, the proposed human feedback reward is applicable only for the object detection tasks because of the IoU component discussed earlier. Therefore, we refrain from performing this experiment on the NER task.

\begin{table}[!h]
\centering
\scalebox{0.95}{
\begin{tabular}{lcc}
\toprule
 & \multicolumn{2}{c}{AP $\longrightarrow$} \\ 
{Method $\downarrow$}   & GROTOAP2 & VOC2007 \\
\midrule
Random                  & 44.127 & 45.541 \\
Entropy Max             & 44.951 & 46.433 \\
Entropy Sum             & 43.842 & 46.437  \\
Margin                  & 42.690 & 45.639 \\
OPAD                    & 45.813 & 46.708 \\
OPAD (Feedback $\lambda$ - 0.1) & \textbf{48.524} & {\textbf{47.238}} \\
OPAD (Feedback $\lambda$ - 0.25) & 46.266 & {46.835} \\
OPAD (Feedback $\lambda$ - 0.40) & 44.899 & 46.646 \\
OPAD (Feedback $\lambda$ - 0.70) & 44.839 & 46.071 \\
OPAD (Feedback $\lambda$ - 1.0) & 44.110 & 46.304 \\
\bottomrule
\end{tabular}}
\caption{Performance of our method with human feedback reward for weak labelling on GROTOAP2 and VOC2007. AP after consuming a total budget of 1152 samples.}
\label{tab:feedback}
\end{table}

\section{Conclusion and Future Works}
We present a robust policy-based method for active learning task in complex content detection problems. The problem of active learning in detection is formulated using a DQN-based sampling network, optimized for performance metrics of the classifier. We extend the active learning setting to weak labelling, and propose rewards for class balance and human feedback. To the best of our knowledge, this is first-of-its-kind work optimizing active learning for detection tasks in documents. We further show the efficacy of the proposed methods on two document analysis tasks and a related object detection task by evaluating our models on 3 large datasets, with significantly better performance. As a future direction, we would like to improve on the DQN, exploiting the recent advances in the field. Currently, our model uses generic features that are common across the different tasks. We would look into more nuanced details that are specific to the documents which would further improve the performance of our models on layout detection and named entity recognition tasks. We would also like to investigate Generative Adversarial Networks or self-supervision for learning from unlabelled data.

\balance
\bibliographystyle{ACM-Reference-Format}
\bibliography{main}


\begin{thebibliography}{61}


\ifx \showCODEN    \undefined \def \showCODEN     #1{\unskip}     \fi
\ifx \showDOI      \undefined \def \showDOI       #1{#1}\fi
\ifx \showISBNx    \undefined \def \showISBNx     #1{\unskip}     \fi
\ifx \showISBNxiii \undefined \def \showISBNxiii  #1{\unskip}     \fi
\ifx \showISSN     \undefined \def \showISSN      #1{\unskip}     \fi
\ifx \showLCCN     \undefined \def \showLCCN      #1{\unskip}     \fi
\ifx \shownote     \undefined \def \shownote      #1{#1}          \fi
\ifx \showarticletitle \undefined \def \showarticletitle #1{#1}   \fi
\ifx \showURL      \undefined \def \showURL       {\relax}        \fi
\providecommand\bibfield[2]{#2}
\providecommand\bibinfo[2]{#2}
\providecommand\natexlab[1]{#1}
\providecommand\showeprint[2][]{arXiv:#2}

\bibitem[\protect\citeauthoryear{Adhikari, Ram, Tang, and Lin}{Adhikari
  et~al\mbox{.}}{2019}]%
        {adhikari2019docbert}
\bibfield{author}{\bibinfo{person}{Ashutosh Adhikari}, \bibinfo{person}{Achyudh
  Ram}, \bibinfo{person}{Raphael Tang}, {and} \bibinfo{person}{Jimmy Lin}.}
  \bibinfo{year}{2019}\natexlab{}.
\newblock \showarticletitle{Docbert: Bert for document classification}.
\newblock \bibinfo{journal}{\emph{arXiv preprint arXiv:1904.08398}}
  (\bibinfo{year}{2019}).
\newblock


\bibitem[\protect\citeauthoryear{Aggarwal, Kong, Gu, Han, and Yu}{Aggarwal
  et~al\mbox{.}}{2014}]%
        {aggarwal2014Active}
\bibfield{author}{\bibinfo{person}{{Charu C.} Aggarwal},
  \bibinfo{person}{Xiangnan Kong}, \bibinfo{person}{Quanquan Gu},
  \bibinfo{person}{Jiawei Han}, {and} \bibinfo{person}{{Philip S.} Yu}.}
  \bibinfo{year}{2014}\natexlab{}.
\newblock \bibinfo{booktitle}{\emph{Active learning: A survey}}.
\newblock \bibinfo{publisher}{CRC Press}, \bibinfo{pages}{571--605}.
\newblock
\showISBNx{9781466586741}
\urldef\tempurl%
\url{https://doi.org/10.1201/b17320}
\showDOI{\tempurl}


\bibitem[\protect\citeauthoryear{Aghdam, Gonzalez-Garcia, Lopez, and
  Weijer}{Aghdam et~al\mbox{.}}{2019}]%
        {aghdam2019PedestrianAL}
\bibfield{author}{\bibinfo{person}{Hamed~H. Aghdam}, \bibinfo{person}{Abel
  Gonzalez-Garcia}, \bibinfo{person}{Antonio Lopez}, {and}
  \bibinfo{person}{Joost Weijer}.} \bibinfo{year}{2019}\natexlab{}.
\newblock \showarticletitle{Active Learning for Deep Detection Neural
  Networks}.
\newblock \bibinfo{journal}{\emph{2019 IEEE/CVF International Conference on
  Computer Vision (ICCV)}} (\bibinfo{date}{Oct} \bibinfo{year}{2019}).
\newblock
\showISBNx{9781728148038}
\urldef\tempurl%
\url{https://doi.org/10.1109/iccv.2019.00377}
\showDOI{\tempurl}


\bibitem[\protect\citeauthoryear{Audebert, Herold, Slimani, and Vidal}{Audebert
  et~al\mbox{.}}{2019}]%
        {audebert2019multimodal}
\bibfield{author}{\bibinfo{person}{Nicolas Audebert},
  \bibinfo{person}{Catherine Herold}, \bibinfo{person}{Kuider Slimani}, {and}
  \bibinfo{person}{C{\'e}dric Vidal}.} \bibinfo{year}{2019}\natexlab{}.
\newblock \showarticletitle{Multimodal deep networks for text and image-based
  document classification}. In \bibinfo{booktitle}{\emph{Joint European
  Conference on Machine Learning and Knowledge Discovery in Databases}}.
  Springer, \bibinfo{pages}{427--443}.
\newblock


\bibitem[\protect\citeauthoryear{Binmakhashen and Mahmoud}{Binmakhashen and
  Mahmoud}{2019}]%
        {binmakhashen2019document}
\bibfield{author}{\bibinfo{person}{Galal~M Binmakhashen} {and}
  \bibinfo{person}{Sabri~A Mahmoud}.} \bibinfo{year}{2019}\natexlab{}.
\newblock \showarticletitle{Document layout analysis: a comprehensive survey}.
\newblock \bibinfo{journal}{\emph{ACM Computing Surveys (CSUR)}}
  \bibinfo{volume}{52}, \bibinfo{number}{6} (\bibinfo{year}{2019}),
  \bibinfo{pages}{1--36}.
\newblock


\bibitem[\protect\citeauthoryear{Bouguelia, Bela{\"\i}d, and
  Bela{\"\i}d}{Bouguelia et~al\mbox{.}}{2013}]%
        {bouguelia2013stream}
\bibfield{author}{\bibinfo{person}{Mohamed-Rafik Bouguelia},
  \bibinfo{person}{Yolande Bela{\"\i}d}, {and} \bibinfo{person}{Abdel
  Bela{\"\i}d}.} \bibinfo{year}{2013}\natexlab{}.
\newblock \showarticletitle{A stream-based semi-supervised active learning
  approach for document classification}. In \bibinfo{booktitle}{\emph{2013 12th
  International Conference on Document Analysis and Recognition}}. IEEE,
  \bibinfo{pages}{611--615}.
\newblock


\bibitem[\protect\citeauthoryear{Brust, Käding, and Denzler}{Brust
  et~al\mbox{.}}{2019}]%
        {brust2019DetectionHeuristicAL}
\bibfield{author}{\bibinfo{person}{Clemens-Alexander Brust},
  \bibinfo{person}{Christoph Käding}, {and} \bibinfo{person}{Joachim
  Denzler}.} \bibinfo{year}{2019}\natexlab{}.
\newblock \showarticletitle{Active Learning for Deep Object Detection}.
\newblock \bibinfo{journal}{\emph{Proceedings of the 14th International Joint
  Conference on Computer Vision, Imaging and Computer Graphics Theory and
  Applications}} (\bibinfo{year}{2019}).
\newblock
\showISBNx{9789897583544}
\urldef\tempurl%
\url{https://doi.org/10.5220/0007248601810190}
\showDOI{\tempurl}


\bibitem[\protect\citeauthoryear{Brust, Käding, and Denzler}{Brust
  et~al\mbox{.}}{2020}]%
        {brust2020ALWeakSupervision}
\bibfield{author}{\bibinfo{person}{Clemens-Alexander Brust},
  \bibinfo{person}{Christoph Käding}, {and} \bibinfo{person}{Joachim
  Denzler}.} \bibinfo{year}{2020}\natexlab{}.
\newblock \showarticletitle{Active and Incremental Learning with Weak
  Supervision}.
\newblock \bibinfo{journal}{\emph{KI - Künstliche Intelligenz}}
  \bibinfo{volume}{34}, \bibinfo{number}{2} (\bibinfo{date}{Jan}
  \bibinfo{year}{2020}), \bibinfo{pages}{165–180}.
\newblock
\showISSN{1610-1987}
\urldef\tempurl%
\url{https://doi.org/10.1007/s13218-020-00631-4}
\showDOI{\tempurl}


\bibitem[\protect\citeauthoryear{Casanova, Pinheiro, Rostamzadeh, and
  Pal}{Casanova et~al\mbox{.}}{2020}]%
        {casanova2020Reinforced}
\bibfield{author}{\bibinfo{person}{Arantxa Casanova}, \bibinfo{person}{Pedro~O.
  Pinheiro}, \bibinfo{person}{Negar Rostamzadeh}, {and}
  \bibinfo{person}{Christopher~J. Pal}.} \bibinfo{year}{2020}\natexlab{}.
\newblock \showarticletitle{Reinforced active learning for image segmentation}.
  In \bibinfo{booktitle}{\emph{International Conference on Learning
  Representations}}.
\newblock
\urldef\tempurl%
\url{https://openreview.net/forum?id=SkgC6TNFvr}
\showURL{%
\tempurl}


\bibitem[\protect\citeauthoryear{Choudhary, Guttikonda, Chowdhury, and
  Learmonth}{Choudhary et~al\mbox{.}}{2020}]%
        {choudhary2020document}
\bibfield{author}{\bibinfo{person}{Sneha Choudhary}, \bibinfo{person}{Haritha
  Guttikonda}, \bibinfo{person}{Dibyendu~Roy Chowdhury}, {and}
  \bibinfo{person}{Gerard~P Learmonth}.} \bibinfo{year}{2020}\natexlab{}.
\newblock \showarticletitle{Document Retrieval Using Deep Learning}. In
  \bibinfo{booktitle}{\emph{2020 Systems and Information Engineering Design
  Symposium (SIEDS)}}. IEEE, \bibinfo{pages}{1--6}.
\newblock


\bibitem[\protect\citeauthoryear{Desai, Lagandula, Guo, Ninomiya, and
  Balasubramanian}{Desai et~al\mbox{.}}{2019}]%
        {desai2019AnAS}
\bibfield{author}{\bibinfo{person}{Sai~Vikas Desai},
  \bibinfo{person}{Akshay~Chandra Lagandula}, \bibinfo{person}{Wei Guo},
  \bibinfo{person}{Seishi Ninomiya}, {and} \bibinfo{person}{Vineeth~N.
  Balasubramanian}.} \bibinfo{year}{2019}\natexlab{}.
\newblock \showarticletitle{An Adaptive Supervision Framework for Active
  Learning in Object Detection}. In \bibinfo{booktitle}{\emph{BMVC}}.
\newblock


\bibitem[\protect\citeauthoryear{Devlin, Chang, Lee, and Toutanova}{Devlin
  et~al\mbox{.}}{2019}]%
        {devlin2019Bert}
\bibfield{author}{\bibinfo{person}{Jacob Devlin}, \bibinfo{person}{Ming-Wei
  Chang}, \bibinfo{person}{Kenton Lee}, {and} \bibinfo{person}{Kristina
  Toutanova}.} \bibinfo{year}{2019}\natexlab{}.
\newblock \showarticletitle{{BERT}: Pre-training of Deep Bidirectional
  Transformers for Language Understanding}. In
  \bibinfo{booktitle}{\emph{Proceedings of the 2019 Conference of the North
  {A}merican Chapter of the Association for Computational Linguistics: Human
  Language Technologies, Volume 1 (Long and Short Papers)}}.
  \bibinfo{publisher}{Association for Computational Linguistics},
  \bibinfo{address}{Minneapolis, Minnesota}, \bibinfo{pages}{4171--4186}.
\newblock
\urldef\tempurl%
\url{https://doi.org/10.18653/v1/N19-1423}
\showDOI{\tempurl}


\bibitem[\protect\citeauthoryear{Ducoffe and Precioso}{Ducoffe and
  Precioso}{2018}]%
        {ducoffe2018Adversarial}
\bibfield{author}{\bibinfo{person}{Melanie Ducoffe} {and}
  \bibinfo{person}{Frederic Precioso}.} \bibinfo{year}{2018}\natexlab{}.
\newblock \bibinfo{title}{Adversarial Active Learning for Deep Networks: a
  Margin Based Approach}.
\newblock
\newblock
\showeprint[arxiv]{1802.09841}~[cs.LG]


\bibitem[\protect\citeauthoryear{Everingham, Van~Gool, Williams, Winn, and
  Zisserman}{Everingham et~al\mbox{.}}{[n.d.]}]%
        {pascal-voc-2007}
\bibfield{author}{\bibinfo{person}{M. Everingham}, \bibinfo{person}{L.
  Van~Gool}, \bibinfo{person}{C.~K.~I. Williams}, \bibinfo{person}{J. Winn},
  {and} \bibinfo{person}{A. Zisserman}.} \bibinfo{year}{[n.d.]}\natexlab{}.
\newblock \bibinfo{title}{The {PASCAL} {V}isual {O}bject {C}lasses {C}hallenge
  2007 {(VOC2007)} {R}esults}.
\newblock
  \bibinfo{howpublished}{http://www.pascal-network.org/challenges/VOC/voc2007/workshop/index.html}.
\newblock


\bibitem[\protect\citeauthoryear{Fang, Li, and Cohn}{Fang
  et~al\mbox{.}}{2017}]%
        {fang2017NlpRLAL}
\bibfield{author}{\bibinfo{person}{Meng Fang}, \bibinfo{person}{Yuan Li}, {and}
  \bibinfo{person}{Trevor Cohn}.} \bibinfo{year}{2017}\natexlab{}.
\newblock \showarticletitle{Learning how to Active Learn: A Deep Reinforcement
  Learning Approach}.
\newblock \bibinfo{journal}{\emph{Proceedings of the 2017 Conference on
  Empirical Methods in Natural Language Processing}} (\bibinfo{year}{2017}).
\newblock
\urldef\tempurl%
\url{https://doi.org/10.18653/v1/d17-1063}
\showDOI{\tempurl}


\bibitem[\protect\citeauthoryear{Freytag, Rodner, and Denzler}{Freytag
  et~al\mbox{.}}{2014}]%
        {freytag2014ALModelOutputChange}
\bibfield{author}{\bibinfo{person}{Alexander Freytag}, \bibinfo{person}{Erik
  Rodner}, {and} \bibinfo{person}{Joachim Denzler}.}
  \bibinfo{year}{2014}\natexlab{}.
\newblock \showarticletitle{Selecting Influential Examples: Active Learning
  with Expected Model Output Changes}. In \bibinfo{booktitle}{\emph{Computer
  Vision -- ECCV 2014}}, \bibfield{editor}{\bibinfo{person}{David Fleet},
  \bibinfo{person}{Tomas Pajdla}, \bibinfo{person}{Bernt Schiele}, {and}
  \bibinfo{person}{Tinne Tuytelaars}} (Eds.). \bibinfo{publisher}{Springer
  International Publishing}, \bibinfo{address}{Cham},
  \bibinfo{pages}{562--577}.
\newblock
\showISBNx{978-3-319-10593-2}


\bibitem[\protect\citeauthoryear{Gal, Islam, and Ghahramani}{Gal
  et~al\mbox{.}}{2017}]%
        {gal2017BayesianAL}
\bibfield{author}{\bibinfo{person}{Yarin Gal}, \bibinfo{person}{Riashat Islam},
  {and} \bibinfo{person}{Zoubin Ghahramani}.} \bibinfo{year}{2017}\natexlab{}.
\newblock \showarticletitle{Deep {B}ayesian Active Learning with Image Data}.
  In \bibinfo{booktitle}{\emph{Proceedings of the 34th International Conference
  on Machine Learning}} \emph{(\bibinfo{series}{Proceedings of Machine Learning
  Research}, Vol.~\bibinfo{volume}{70})},
  \bibfield{editor}{\bibinfo{person}{Doina Precup} {and}
  \bibinfo{person}{Yee~Whye Teh}} (Eds.). \bibinfo{publisher}{PMLR},
  \bibinfo{address}{International Convention Centre, Sydney, Australia},
  \bibinfo{pages}{1183--1192}.
\newblock
\urldef\tempurl%
\url{http://proceedings.mlr.press/v70/gal17a.html}
\showURL{%
\tempurl}


\bibitem[\protect\citeauthoryear{Godbole, Harpale, Sarawagi, and
  Chakrabarti}{Godbole et~al\mbox{.}}{2004}]%
        {godbole2004document}
\bibfield{author}{\bibinfo{person}{Shantanu Godbole}, \bibinfo{person}{Abhay
  Harpale}, \bibinfo{person}{Sunita Sarawagi}, {and} \bibinfo{person}{Soumen
  Chakrabarti}.} \bibinfo{year}{2004}\natexlab{}.
\newblock \showarticletitle{Document classification through interactive
  supervision of document and term labels}. In
  \bibinfo{booktitle}{\emph{European Conference on Principles of Data Mining
  and Knowledge Discovery}}. Springer, \bibinfo{pages}{185--196}.
\newblock


\bibitem[\protect\citeauthoryear{Gr{\"u}ning, Leifert, Strau{\ss}, Michael, and
  Labahn}{Gr{\"u}ning et~al\mbox{.}}{2019}]%
        {gruning2019two}
\bibfield{author}{\bibinfo{person}{Tobias Gr{\"u}ning},
  \bibinfo{person}{Gundram Leifert}, \bibinfo{person}{Tobias Strau{\ss}},
  \bibinfo{person}{Johannes Michael}, {and} \bibinfo{person}{Roger Labahn}.}
  \bibinfo{year}{2019}\natexlab{}.
\newblock \showarticletitle{A two-stage method for text line detection in
  historical documents}.
\newblock \bibinfo{journal}{\emph{International Journal on Document Analysis
  and Recognition (IJDAR)}} \bibinfo{volume}{22}, \bibinfo{number}{3}
  (\bibinfo{year}{2019}), \bibinfo{pages}{285--302}.
\newblock


\bibitem[\protect\citeauthoryear{Guo and Schuurmans}{Guo and
  Schuurmans}{2007}]%
        {guo2007discriminative}
\bibfield{author}{\bibinfo{person}{Yuhong Guo} {and} \bibinfo{person}{Dale
  Schuurmans}.} \bibinfo{year}{2007}\natexlab{}.
\newblock \showarticletitle{Discriminative Batch Mode Active Learning.}. In
  \bibinfo{booktitle}{\emph{NIPS}}. Citeseer, \bibinfo{pages}{593--600}.
\newblock


\bibitem[\protect\citeauthoryear{Harley, Ufkes, and Derpanis}{Harley
  et~al\mbox{.}}{2015}]%
        {harley2015evaluation}
\bibfield{author}{\bibinfo{person}{Adam~W Harley}, \bibinfo{person}{Alex
  Ufkes}, {and} \bibinfo{person}{Konstantinos~G Derpanis}.}
  \bibinfo{year}{2015}\natexlab{}.
\newblock \showarticletitle{Evaluation of deep convolutional nets for document
  image classification and retrieval}. In \bibinfo{booktitle}{\emph{2015 13th
  International Conference on Document Analysis and Recognition (ICDAR)}}.
  IEEE, \bibinfo{pages}{991--995}.
\newblock


\bibitem[\protect\citeauthoryear{Haussmann, Hamprecht, and Kandemir}{Haussmann
  et~al\mbox{.}}{2019}]%
        {haussmann2019CLassRLAL}
\bibfield{author}{\bibinfo{person}{Manuel Haussmann}, \bibinfo{person}{Fred
  Hamprecht}, {and} \bibinfo{person}{Melih Kandemir}.}
  \bibinfo{year}{2019}\natexlab{}.
\newblock \showarticletitle{Deep Active Learning with Adaptive Acquisition}.
\newblock \bibinfo{journal}{\emph{Proceedings of the Twenty-Eighth
  International Joint Conference on Artificial Intelligence}}
  (\bibinfo{date}{Aug} \bibinfo{year}{2019}).
\newblock
\showISBNx{9780999241141}
\urldef\tempurl%
\url{https://doi.org/10.24963/ijcai.2019/343}
\showDOI{\tempurl}


\bibitem[\protect\citeauthoryear{He, Zhang, Ren, and Sun}{He
  et~al\mbox{.}}{2015}]%
        {he2015Deep}
\bibfield{author}{\bibinfo{person}{Kaiming He}, \bibinfo{person}{Xiangyu
  Zhang}, \bibinfo{person}{Shaoqing Ren}, {and} \bibinfo{person}{Jian Sun}.}
  \bibinfo{year}{2015}\natexlab{}.
\newblock \bibinfo{title}{Deep Residual Learning for Image Recognition}.
\newblock
\newblock
\showeprint[arxiv]{1512.03385}~[cs.CV]


\bibitem[\protect\citeauthoryear{Houlsby, Huszar, Ghahramani, and
  Hern\'{a}ndez-lobato}{Houlsby et~al\mbox{.}}{2012}]%
        {houlsby2012BALD}
\bibfield{author}{\bibinfo{person}{Neil Houlsby}, \bibinfo{person}{Ferenc
  Huszar}, \bibinfo{person}{Zoubin Ghahramani}, {and} \bibinfo{person}{Jose~M.
  Hern\'{a}ndez-lobato}.} \bibinfo{year}{2012}\natexlab{}.
\newblock \showarticletitle{Collaborative Gaussian Processes for Preference
  Learning}.
\newblock In \bibinfo{booktitle}{\emph{Advances in Neural Information
  Processing Systems 25}}, \bibfield{editor}{\bibinfo{person}{F.~Pereira},
  \bibinfo{person}{C.~J.~C. Burges}, \bibinfo{person}{L.~Bottou}, {and}
  \bibinfo{person}{K.~Q. Weinberger}} (Eds.). \bibinfo{publisher}{Curran
  Associates, Inc.}, \bibinfo{pages}{2096--2104}.
\newblock
\urldef\tempurl%
\url{http://papers.nips.cc/paper/4700-collaborative-gaussian-processes-for-preference-learning.pdf}
\showURL{%
\tempurl}


\bibitem[\protect\citeauthoryear{Huang, Xu, and Yu}{Huang
  et~al\mbox{.}}{2015}]%
        {huang2015bidirectional}
\bibfield{author}{\bibinfo{person}{Zhiheng Huang}, \bibinfo{person}{Wei Xu},
  {and} \bibinfo{person}{Kai Yu}.} \bibinfo{year}{2015}\natexlab{}.
\newblock \showarticletitle{Bidirectional LSTM-CRF models for sequence
  tagging}.
\newblock \bibinfo{journal}{\emph{arXiv preprint arXiv:1508.01991}}
  (\bibinfo{year}{2015}).
\newblock


\bibitem[\protect\citeauthoryear{Jain and Wigington}{Jain and
  Wigington}{2019}]%
        {jain2019multimodal}
\bibfield{author}{\bibinfo{person}{Rajiv Jain} {and} \bibinfo{person}{Curtis
  Wigington}.} \bibinfo{year}{2019}\natexlab{}.
\newblock \showarticletitle{Multimodal Document Image Classification}. In
  \bibinfo{booktitle}{\emph{2019 International Conference on Document Analysis
  and Recognition (ICDAR)}}. IEEE, \bibinfo{pages}{71--77}.
\newblock


\bibitem[\protect\citeauthoryear{Lewis, Agam, Argamon, Frieder, Grossman, and
  Heard}{Lewis et~al\mbox{.}}{2006}]%
        {lewis2006building}
\bibfield{author}{\bibinfo{person}{David Lewis}, \bibinfo{person}{Gady Agam},
  \bibinfo{person}{Shlomo Argamon}, \bibinfo{person}{Ophir Frieder},
  \bibinfo{person}{David Grossman}, {and} \bibinfo{person}{Jefferson Heard}.}
  \bibinfo{year}{2006}\natexlab{}.
\newblock \showarticletitle{Building a test collection for complex document
  information processing}. In \bibinfo{booktitle}{\emph{Proceedings of the 29th
  annual international ACM SIGIR conference on Research and development in
  information retrieval}}. \bibinfo{pages}{665--666}.
\newblock


\bibitem[\protect\citeauthoryear{Li, Xu, Cui, Huang, Wei, Li, and Zhou}{Li
  et~al\mbox{.}}{2020}]%
        {li2020docbank}
\bibfield{author}{\bibinfo{person}{Minghao Li}, \bibinfo{person}{Yiheng Xu},
  \bibinfo{person}{Lei Cui}, \bibinfo{person}{Shaohan Huang},
  \bibinfo{person}{Furu Wei}, \bibinfo{person}{Zhoujun Li}, {and}
  \bibinfo{person}{Ming Zhou}.} \bibinfo{year}{2020}\natexlab{}.
\newblock \showarticletitle{Docbank: A benchmark dataset for document layout
  analysis}.
\newblock \bibinfo{journal}{\emph{arXiv preprint arXiv:2006.01038}}
  (\bibinfo{year}{2020}).
\newblock


\bibitem[\protect\citeauthoryear{Lin, Maire, Belongie, Hays, Perona, Ramanan,
  Doll{\'a}r, and Zitnick}{Lin et~al\mbox{.}}{2014}]%
        {lin2014microsoft}
\bibfield{author}{\bibinfo{person}{Tsung-Yi Lin}, \bibinfo{person}{Michael
  Maire}, \bibinfo{person}{Serge Belongie}, \bibinfo{person}{James Hays},
  \bibinfo{person}{Pietro Perona}, \bibinfo{person}{Deva Ramanan},
  \bibinfo{person}{Piotr Doll{\'a}r}, {and} \bibinfo{person}{C~Lawrence
  Zitnick}.} \bibinfo{year}{2014}\natexlab{}.
\newblock \showarticletitle{Microsoft coco: Common objects in context}. In
  \bibinfo{booktitle}{\emph{European conference on computer vision}}. Springer,
  \bibinfo{pages}{740--755}.
\newblock


\bibitem[\protect\citeauthoryear{Liu, Buntine, and Haffari}{Liu
  et~al\mbox{.}}{2018a}]%
        {liu2018ImmitationAL}
\bibfield{author}{\bibinfo{person}{Ming Liu}, \bibinfo{person}{Wray Buntine},
  {and} \bibinfo{person}{Gholamreza Haffari}.}
  \bibinfo{year}{2018}\natexlab{a}.
\newblock \showarticletitle{Learning How to Actively Learn: A Deep Imitation
  Learning Approach}. In \bibinfo{booktitle}{\emph{Proceedings of the 56th
  Annual Meeting of the Association for Computational Linguistics (Volume 1:
  Long Papers)}}. \bibinfo{publisher}{Association for Computational
  Linguistics}, \bibinfo{address}{Melbourne, Australia},
  \bibinfo{pages}{1874--1883}.
\newblock
\urldef\tempurl%
\url{https://doi.org/10.18653/v1/P18-1174}
\showDOI{\tempurl}


\bibitem[\protect\citeauthoryear{Liu, Buntine, and Haffari}{Liu
  et~al\mbox{.}}{2018b}]%
        {liu2018MtRLAL}
\bibfield{author}{\bibinfo{person}{Ming Liu}, \bibinfo{person}{Wray Buntine},
  {and} \bibinfo{person}{Gholamreza Haffari}.}
  \bibinfo{year}{2018}\natexlab{b}.
\newblock \showarticletitle{Learning to Actively Learn Neural Machine
  Translation}. In \bibinfo{booktitle}{\emph{Proceedings of the 22nd Conference
  on Computational Natural Language Learning}}. \bibinfo{publisher}{Association
  for Computational Linguistics}, \bibinfo{address}{Brussels, Belgium},
  \bibinfo{pages}{334--344}.
\newblock
\urldef\tempurl%
\url{https://doi.org/10.18653/v1/K18-1033}
\showDOI{\tempurl}


\bibitem[\protect\citeauthoryear{{Liu}, {Wang}, {Gong}, {Tao}, and {Lu}}{{Liu}
  et~al\mbox{.}}{2019}]%
        {liu2019PersonReIdAL}
\bibfield{author}{\bibinfo{person}{Z. {Liu}}, \bibinfo{person}{J. {Wang}},
  \bibinfo{person}{S. {Gong}}, \bibinfo{person}{D. {Tao}}, {and}
  \bibinfo{person}{H. {Lu}}.} \bibinfo{year}{2019}\natexlab{}.
\newblock \showarticletitle{Deep Reinforcement Active Learning for
  Human-in-the-Loop Person Re-Identification}. In
  \bibinfo{booktitle}{\emph{2019 IEEE/CVF International Conference on Computer
  Vision (ICCV)}}. \bibinfo{pages}{6121--6130}.
\newblock


\bibitem[\protect\citeauthoryear{Luo, Yang, Yang, Zhang, Wang, Lin, and
  Wang}{Luo et~al\mbox{.}}{2018}]%
        {luo2018attention}
\bibfield{author}{\bibinfo{person}{Ling Luo}, \bibinfo{person}{Zhihao Yang},
  \bibinfo{person}{Pei Yang}, \bibinfo{person}{Yin Zhang}, \bibinfo{person}{Lei
  Wang}, \bibinfo{person}{Hongfei Lin}, {and} \bibinfo{person}{Jian Wang}.}
  \bibinfo{year}{2018}\natexlab{}.
\newblock \showarticletitle{An attention-based BiLSTM-CRF approach to
  document-level chemical named entity recognition}.
\newblock \bibinfo{journal}{\emph{Bioinformatics}} \bibinfo{volume}{34},
  \bibinfo{number}{8} (\bibinfo{year}{2018}), \bibinfo{pages}{1381--1388}.
\newblock


\bibitem[\protect\citeauthoryear{Mayer and Timofte}{Mayer and Timofte}{2020}]%
        {Mayer2020AdversarialSF}
\bibfield{author}{\bibinfo{person}{Christoph Mayer} {and} \bibinfo{person}{Radu
  Timofte}.} \bibinfo{year}{2020}\natexlab{}.
\newblock \showarticletitle{Adversarial Sampling for Active Learning}.
\newblock \bibinfo{journal}{\emph{2020 IEEE Winter Conference on Applications
  of Computer Vision (WACV)}} (\bibinfo{year}{2020}),
  \bibinfo{pages}{3060--3068}.
\newblock


\bibitem[\protect\citeauthoryear{Mnih, Kavukcuoglu, Silver, Graves, Antonoglou,
  Wierstra, and Riedmiller}{Mnih et~al\mbox{.}}{2013}]%
        {mnih2013DeepQLearning}
\bibfield{author}{\bibinfo{person}{Volodymyr Mnih}, \bibinfo{person}{Koray
  Kavukcuoglu}, \bibinfo{person}{David Silver}, \bibinfo{person}{Alex Graves},
  \bibinfo{person}{Ioannis Antonoglou}, \bibinfo{person}{Daan Wierstra}, {and}
  \bibinfo{person}{Martin~A. Riedmiller}.} \bibinfo{year}{2013}\natexlab{}.
\newblock \showarticletitle{Playing Atari with Deep Reinforcement Learning}.
\newblock \bibinfo{journal}{\emph{ArXiv}}  \bibinfo{volume}{abs/1312.5602}
  (\bibinfo{year}{2013}).
\newblock


\bibitem[\protect\citeauthoryear{Oliveira, Seguin, and Kaplan}{Oliveira
  et~al\mbox{.}}{2018}]%
        {oliveira2018dhsegment}
\bibfield{author}{\bibinfo{person}{Sofia~Ares Oliveira},
  \bibinfo{person}{Benoit Seguin}, {and} \bibinfo{person}{Frederic Kaplan}.}
  \bibinfo{year}{2018}\natexlab{}.
\newblock \showarticletitle{dhSegment: A generic deep-learning approach for
  document segmentation}. In \bibinfo{booktitle}{\emph{2018 16th International
  Conference on Frontiers in Handwriting Recognition (ICFHR)}}. IEEE,
  \bibinfo{pages}{7--12}.
\newblock


\bibitem[\protect\citeauthoryear{{Papadopoulos}, {Uijlings}, {Keller}, and
  {Ferrari}}{{Papadopoulos} et~al\mbox{.}}{2016}]%
        {papadopoulos2016NoBBoxesHumanVerification}
\bibfield{author}{\bibinfo{person}{D.~P. {Papadopoulos}},
  \bibinfo{person}{J.~R.~R. {Uijlings}}, \bibinfo{person}{F. {Keller}}, {and}
  \bibinfo{person}{V. {Ferrari}}.} \bibinfo{year}{2016}\natexlab{}.
\newblock \showarticletitle{We Don’t Need No Bounding-Boxes: Training Object
  Class Detectors Using Only Human Verification}. In
  \bibinfo{booktitle}{\emph{2016 IEEE Conference on Computer Vision and Pattern
  Recognition (CVPR)}}. \bibinfo{pages}{854--863}.
\newblock


\bibitem[\protect\citeauthoryear{Papadopoulos, Uijlings, Keller, and
  Ferrari}{Papadopoulos et~al\mbox{.}}{2017}]%
        {papadopoulos2017ClickSupervision}
\bibfield{author}{\bibinfo{person}{Dim~P. Papadopoulos},
  \bibinfo{person}{Jasper R.~R. Uijlings}, \bibinfo{person}{Frank Keller},
  {and} \bibinfo{person}{Vittorio Ferrari}.} \bibinfo{year}{2017}\natexlab{}.
\newblock \showarticletitle{Training Object Class Detectors with Click
  Supervision}.
\newblock \bibinfo{journal}{\emph{2017 IEEE Conference on Computer Vision and
  Pattern Recognition (CVPR)}} (\bibinfo{date}{Jul} \bibinfo{year}{2017}).
\newblock
\showISBNx{9781538604571}
\urldef\tempurl%
\url{https://doi.org/10.1109/cvpr.2017.27}
\showDOI{\tempurl}


\bibitem[\protect\citeauthoryear{Pappagari, Zelasko, Villalba, Carmiel, and
  Dehak}{Pappagari et~al\mbox{.}}{2019}]%
        {pappagari2019hierarchical}
\bibfield{author}{\bibinfo{person}{Raghavendra Pappagari},
  \bibinfo{person}{Piotr Zelasko}, \bibinfo{person}{Jes{\'u}s Villalba},
  \bibinfo{person}{Yishay Carmiel}, {and} \bibinfo{person}{Najim Dehak}.}
  \bibinfo{year}{2019}\natexlab{}.
\newblock \showarticletitle{Hierarchical transformers for long document
  classification}. In \bibinfo{booktitle}{\emph{2019 IEEE Automatic Speech
  Recognition and Understanding Workshop (ASRU)}}. IEEE,
  \bibinfo{pages}{838--844}.
\newblock


\bibitem[\protect\citeauthoryear{Ramshaw and Marcus}{Ramshaw and
  Marcus}{1999}]%
        {ramshaw1999text}
\bibfield{author}{\bibinfo{person}{Lance~A Ramshaw} {and}
  \bibinfo{person}{Mitchell~P Marcus}.} \bibinfo{year}{1999}\natexlab{}.
\newblock \showarticletitle{Text chunking using transformation-based learning}.
\newblock In \bibinfo{booktitle}{\emph{Natural language processing using very
  large corpora}}. \bibinfo{publisher}{Springer}, \bibinfo{pages}{157--176}.
\newblock


\bibitem[\protect\citeauthoryear{Ren, He, Girshick, and Sun}{Ren
  et~al\mbox{.}}{2015a}]%
        {ren2015FasterRcnn}
\bibfield{author}{\bibinfo{person}{Shaoqing Ren}, \bibinfo{person}{Kaiming He},
  \bibinfo{person}{Ross Girshick}, {and} \bibinfo{person}{Jian Sun}.}
  \bibinfo{year}{2015}\natexlab{a}.
\newblock \showarticletitle{Faster R-CNN: Towards Real-Time Object Detection
  with Region Proposal Networks}.
\newblock In \bibinfo{booktitle}{\emph{Advances in Neural Information
  Processing Systems 28}}, \bibfield{editor}{\bibinfo{person}{C.~Cortes},
  \bibinfo{person}{N.~D. Lawrence}, \bibinfo{person}{D.~D. Lee},
  \bibinfo{person}{M.~Sugiyama}, {and} \bibinfo{person}{R.~Garnett}} (Eds.).
  \bibinfo{publisher}{Curran Associates, Inc.}, \bibinfo{pages}{91--99}.
\newblock
\urldef\tempurl%
\url{http://papers.nips.cc/paper/5638-faster-r-cnn-towards-real-time-object-detection-with-region-proposal-networks.pdf}
\showURL{%
\tempurl}


\bibitem[\protect\citeauthoryear{Ren, He, Girshick, and Sun}{Ren
  et~al\mbox{.}}{2015b}]%
        {ren2015faster}
\bibfield{author}{\bibinfo{person}{Shaoqing Ren}, \bibinfo{person}{Kaiming He},
  \bibinfo{person}{Ross Girshick}, {and} \bibinfo{person}{Jian Sun}.}
  \bibinfo{year}{2015}\natexlab{b}.
\newblock \showarticletitle{Faster r-cnn: Towards real-time object detection
  with region proposal networks}. In \bibinfo{booktitle}{\emph{Advances in
  neural information processing systems}}. \bibinfo{pages}{91--99}.
\newblock


\bibitem[\protect\citeauthoryear{Roy, Unmesh, and Namboodiri}{Roy
  et~al\mbox{.}}{2018}]%
        {roy2018DetectionHeuristicAL}
\bibfield{author}{\bibinfo{person}{Soumya Roy}, \bibinfo{person}{Asim Unmesh},
  {and} \bibinfo{person}{Vinay~P Namboodiri}.} \bibinfo{year}{2018}\natexlab{}.
\newblock \showarticletitle{Deep active learning for object detection.}. In
  \bibinfo{booktitle}{\emph{BMVC}}. \bibinfo{pages}{91}.
\newblock


\bibitem[\protect\citeauthoryear{Settles}{Settles}{2009}]%
        {settles2009Active}
\bibfield{author}{\bibinfo{person}{Burr Settles}.}
  \bibinfo{year}{2009}\natexlab{}.
\newblock \bibinfo{booktitle}{\emph{Active learning literature survey}}.
\newblock \bibinfo{type}{{T}echnical {R}eport}.
  \bibinfo{institution}{University of Wisconsin-Madison Department of Computer
  Sciences}.
\newblock


\bibitem[\protect\citeauthoryear{Shannon}{Shannon}{2001}]%
        {shannon2001mathematical}
\bibfield{author}{\bibinfo{person}{Claude~Elwood Shannon}.}
  \bibinfo{year}{2001}\natexlab{}.
\newblock \showarticletitle{A mathematical theory of communication}.
\newblock \bibinfo{journal}{\emph{ACM SIGMOBILE mobile computing and
  communications review}} \bibinfo{volume}{5}, \bibinfo{number}{1}
  (\bibinfo{year}{2001}), \bibinfo{pages}{3--55}.
\newblock


\bibitem[\protect\citeauthoryear{Shen, Yun, Lipton, Kronrod, and
  Anandkumar}{Shen et~al\mbox{.}}{2017}]%
        {shen2017NERHeuristicAL}
\bibfield{author}{\bibinfo{person}{Yanyao Shen}, \bibinfo{person}{Hyokun Yun},
  \bibinfo{person}{Zachary Lipton}, \bibinfo{person}{Yakov Kronrod}, {and}
  \bibinfo{person}{Animashree Anandkumar}.} \bibinfo{year}{2017}\natexlab{}.
\newblock \showarticletitle{Deep Active Learning for Named Entity Recognition}.
\newblock \bibinfo{journal}{\emph{Proceedings of the 2nd Workshop on
  Representation Learning for NLP}} (\bibinfo{year}{2017}).
\newblock
\urldef\tempurl%
\url{https://doi.org/10.18653/v1/w17-2630}
\showDOI{\tempurl}


\bibitem[\protect\citeauthoryear{Sugathadasa, Ayesha, de~Silva, Perera,
  Jayawardana, Lakmal, and Perera}{Sugathadasa et~al\mbox{.}}{2018}]%
        {sugathadasa2018legal}
\bibfield{author}{\bibinfo{person}{Keet Sugathadasa}, \bibinfo{person}{Buddhi
  Ayesha}, \bibinfo{person}{Nisansa de Silva}, \bibinfo{person}{Amal~Shehan
  Perera}, \bibinfo{person}{Vindula Jayawardana}, \bibinfo{person}{Dimuthu
  Lakmal}, {and} \bibinfo{person}{Madhavi Perera}.}
  \bibinfo{year}{2018}\natexlab{}.
\newblock \showarticletitle{Legal document retrieval using document vector
  embeddings and deep learning}. In \bibinfo{booktitle}{\emph{Science and
  information conference}}. Springer, \bibinfo{pages}{160--175}.
\newblock


\bibitem[\protect\citeauthoryear{Sutton}{Sutton}{1988}]%
        {sutton1988learning}
\bibfield{author}{\bibinfo{person}{Richard~S Sutton}.}
  \bibinfo{year}{1988}\natexlab{}.
\newblock \showarticletitle{Learning to predict by the methods of temporal
  differences}.
\newblock \bibinfo{journal}{\emph{Machine learning}} \bibinfo{volume}{3},
  \bibinfo{number}{1} (\bibinfo{year}{1988}), \bibinfo{pages}{9--44}.
\newblock


\bibitem[\protect\citeauthoryear{Tjong Kim~Sang and De~Meulder}{Tjong Kim~Sang
  and De~Meulder}{2003}]%
        {tjong2003introduction}
\bibfield{author}{\bibinfo{person}{Erik~F Tjong Kim~Sang} {and}
  \bibinfo{person}{Fien De~Meulder}.} \bibinfo{year}{2003}\natexlab{}.
\newblock \showarticletitle{Introduction to the CoNLL-2003 shared task:
  language-independent named entity recognition}. In
  \bibinfo{booktitle}{\emph{Proceedings of the seventh conference on Natural
  language learning at HLT-NAACL 2003-Volume 4}}. \bibinfo{pages}{142--147}.
\newblock


\bibitem[\protect\citeauthoryear{Tkaczyk, Szostek, and Bolikowski}{Tkaczyk
  et~al\mbox{.}}{2014}]%
        {Tkaczyk2014GROTOAP2T}
\bibfield{author}{\bibinfo{person}{Dominika Tkaczyk}, \bibinfo{person}{Pawel
  Szostek}, {and} \bibinfo{person}{Lukasz Bolikowski}.}
  \bibinfo{year}{2014}\natexlab{}.
\newblock \showarticletitle{GROTOAP2 - The Methodology of Creating a Large
  Ground Truth Dataset of Scientific Articles}.
\newblock \bibinfo{journal}{\emph{D-Lib Mag.}}  \bibinfo{volume}{20}
  (\bibinfo{year}{2014}).
\newblock


\bibitem[\protect\citeauthoryear{Trabelsi, Chen, Davison, and Heflin}{Trabelsi
  et~al\mbox{.}}{2021}]%
        {trabelsi2021neural}
\bibfield{author}{\bibinfo{person}{Mohamed Trabelsi}, \bibinfo{person}{Zhiyu
  Chen}, \bibinfo{person}{Brian~D Davison}, {and} \bibinfo{person}{Jeff
  Heflin}.} \bibinfo{year}{2021}\natexlab{}.
\newblock \showarticletitle{Neural Ranking Models for Document Retrieval}.
\newblock \bibinfo{journal}{\emph{arXiv preprint arXiv:2102.11903}}
  (\bibinfo{year}{2021}).
\newblock


\bibitem[\protect\citeauthoryear{{Wang} and {Shang}}{{Wang} and
  {Shang}}{2014}]%
        {wang2014Heuristic}
\bibfield{author}{\bibinfo{person}{D. {Wang}} {and} \bibinfo{person}{Y.
  {Shang}}.} \bibinfo{year}{2014}\natexlab{}.
\newblock \showarticletitle{A new active labeling method for deep learning}. In
  \bibinfo{booktitle}{\emph{2014 International Joint Conference on Neural
  Networks (IJCNN)}}. \bibinfo{pages}{112--119}.
\newblock


\bibitem[\protect\citeauthoryear{Wang, Zhang, Li, Zhang, and Lin}{Wang
  et~al\mbox{.}}{2017}]%
        {wang2017CEAL}
\bibfield{author}{\bibinfo{person}{Keze Wang}, \bibinfo{person}{Dongyu Zhang},
  \bibinfo{person}{Ya Li}, \bibinfo{person}{Ruimao Zhang}, {and}
  \bibinfo{person}{Liang Lin}.} \bibinfo{year}{2017}\natexlab{}.
\newblock \showarticletitle{Cost-Effective Active Learning for Deep Image
  Classification}.
\newblock \bibinfo{journal}{\emph{IEEE Transactions on Circuits and Systems for
  Video Technology}} \bibinfo{volume}{27}, \bibinfo{number}{12}
  (\bibinfo{date}{Dec} \bibinfo{year}{2017}), \bibinfo{pages}{2591–2600}.
\newblock
\showISSN{1558-2205}
\urldef\tempurl%
\url{https://doi.org/10.1109/tcsvt.2016.2589879}
\showDOI{\tempurl}


\bibitem[\protect\citeauthoryear{Wang, Yao, Kwok, and Ni}{Wang
  et~al\mbox{.}}{2020}]%
        {wang2020generalizing}
\bibfield{author}{\bibinfo{person}{Yaqing Wang}, \bibinfo{person}{Quanming
  Yao}, \bibinfo{person}{James~T Kwok}, {and} \bibinfo{person}{Lionel~M Ni}.}
  \bibinfo{year}{2020}\natexlab{}.
\newblock \showarticletitle{Generalizing from a few examples: A survey on
  few-shot learning}.
\newblock \bibinfo{journal}{\emph{ACM Computing Surveys (CSUR)}}
  \bibinfo{volume}{53}, \bibinfo{number}{3} (\bibinfo{year}{2020}),
  \bibinfo{pages}{1--34}.
\newblock


\bibitem[\protect\citeauthoryear{Wilson and Cook}{Wilson and Cook}{2020}]%
        {wilson2020survey}
\bibfield{author}{\bibinfo{person}{Garrett Wilson} {and}
  \bibinfo{person}{Diane~J Cook}.} \bibinfo{year}{2020}\natexlab{}.
\newblock \showarticletitle{A survey of unsupervised deep domain adaptation}.
\newblock \bibinfo{journal}{\emph{ACM Transactions on Intelligent Systems and
  Technology (TIST)}} \bibinfo{volume}{11}, \bibinfo{number}{5}
  (\bibinfo{year}{2020}), \bibinfo{pages}{1--46}.
\newblock


\bibitem[\protect\citeauthoryear{Xu, Li, Cui, Huang, Wei, and Zhou}{Xu
  et~al\mbox{.}}{2020a}]%
        {xu2020layoutlm}
\bibfield{author}{\bibinfo{person}{Yiheng Xu}, \bibinfo{person}{Minghao Li},
  \bibinfo{person}{Lei Cui}, \bibinfo{person}{Shaohan Huang},
  \bibinfo{person}{Furu Wei}, {and} \bibinfo{person}{Ming Zhou}.}
  \bibinfo{year}{2020}\natexlab{a}.
\newblock \showarticletitle{Layoutlm: Pre-training of text and layout for
  document image understanding}. In \bibinfo{booktitle}{\emph{Proceedings of
  the 26th ACM SIGKDD International Conference on Knowledge Discovery \& Data
  Mining}}. \bibinfo{pages}{1192--1200}.
\newblock


\bibitem[\protect\citeauthoryear{Xu, Xu, Lv, Cui, Wei, Wang, Lu, Florencio,
  Zhang, Che, et~al\mbox{.}}{Xu et~al\mbox{.}}{2020b}]%
        {xu2020layoutlmv2}
\bibfield{author}{\bibinfo{person}{Yang Xu}, \bibinfo{person}{Yiheng Xu},
  \bibinfo{person}{Tengchao Lv}, \bibinfo{person}{Lei Cui},
  \bibinfo{person}{Furu Wei}, \bibinfo{person}{Guoxin Wang},
  \bibinfo{person}{Yijuan Lu}, \bibinfo{person}{Dinei Florencio},
  \bibinfo{person}{Cha Zhang}, \bibinfo{person}{Wanxiang Che}, {et~al\mbox{.}}}
  \bibinfo{year}{2020}\natexlab{b}.
\newblock \showarticletitle{LayoutLMv2: Multi-modal Pre-training for
  Visually-Rich Document Understanding}.
\newblock \bibinfo{journal}{\emph{arXiv preprint arXiv:2012.14740}}
  (\bibinfo{year}{2020}).
\newblock


\bibitem[\protect\citeauthoryear{Yang and Mitchell}{Yang and Mitchell}{2016}]%
        {yang2016joint}
\bibfield{author}{\bibinfo{person}{Bishan Yang} {and} \bibinfo{person}{Tom
  Mitchell}.} \bibinfo{year}{2016}\natexlab{}.
\newblock \showarticletitle{Joint extraction of events and entities within a
  document context}.
\newblock \bibinfo{journal}{\emph{arXiv preprint arXiv:1609.03632}}
  (\bibinfo{year}{2016}).
\newblock


\bibitem[\protect\citeauthoryear{Yang, Song, King, and Xu}{Yang
  et~al\mbox{.}}{2021}]%
        {yang2021survey}
\bibfield{author}{\bibinfo{person}{Xiangli Yang}, \bibinfo{person}{Zixing
  Song}, \bibinfo{person}{Irwin King}, {and} \bibinfo{person}{Zenglin Xu}.}
  \bibinfo{year}{2021}\natexlab{}.
\newblock \showarticletitle{A Survey on Deep Semi-supervised Learning}.
\newblock \bibinfo{journal}{\emph{arXiv preprint arXiv:2103.00550}}
  (\bibinfo{year}{2021}).
\newblock


\bibitem[\protect\citeauthoryear{Yoo and Kweon}{Yoo and Kweon}{2019}]%
        {yoo2019LearningLossAL}
\bibfield{author}{\bibinfo{person}{Donggeun Yoo} {and} \bibinfo{person}{In~So
  Kweon}.} \bibinfo{year}{2019}\natexlab{}.
\newblock \showarticletitle{Learning Loss for Active Learning}.
\newblock \bibinfo{journal}{\emph{2019 IEEE/CVF Conference on Computer Vision
  and Pattern Recognition (CVPR)}} (\bibinfo{date}{Jun} \bibinfo{year}{2019}).
\newblock
\showISBNx{9781728132938}
\urldef\tempurl%
\url{https://doi.org/10.1109/cvpr.2019.00018}
\showDOI{\tempurl}


\bibitem[\protect\citeauthoryear{Zhong, Tang, and Yepes}{Zhong
  et~al\mbox{.}}{2019}]%
        {zhong2019publaynet}
\bibfield{author}{\bibinfo{person}{Xu Zhong}, \bibinfo{person}{Jianbin Tang},
  {and} \bibinfo{person}{Antonio~Jimeno Yepes}.}
  \bibinfo{year}{2019}\natexlab{}.
\newblock \showarticletitle{Publaynet: largest dataset ever for document layout
  analysis}. In \bibinfo{booktitle}{\emph{2019 International Conference on
  Document Analysis and Recognition (ICDAR)}}. IEEE,
  \bibinfo{pages}{1015--1022}.
\newblock


\end{thebibliography}

\end{document}